\pdfoutput=1
\documentclass[conference, 10pt]{IEEEtran}
\IEEEoverridecommandlockouts
\usepackage{cite}
\usepackage{amsmath,amssymb,amsfonts}
\usepackage{graphicx}
\usepackage{textcomp}
\usepackage{xcolor}
\usepackage{bm}
\usepackage{cases}
\usepackage[skip=1pt, belowskip=0pt]{caption}
\usepackage{subcaption} 
\usepackage[ruled,vlined,linesnumbered]{algorithm2e} 
    \usepackage{setspace} 
\usepackage[makeroom]{cancel} 
\usepackage{amsthm}
\usepackage{enumitem}  

\usepackage[normalem]{ulem} 

    \usepackage{physics}
    \usepackage{amsmath}
    \usepackage{tikz}
    \usepackage{mathdots}
    \usepackage{yhmath}
    \usepackage{cancel}
    \usepackage{color}
    \usepackage{siunitx}
    \usepackage{array}
    \usepackage{multirow}
    \usepackage{amssymb}
    \usepackage{gensymb}
    \usepackage{tabularx}
    \usepackage{booktabs}
    \usetikzlibrary{patterns}
    \usetikzlibrary{shapes}
    
\makeatletter
\def\BibTeX{{\rm B\kern-.05em{\sc i\kern-.025em b}\kern-.08em
    T\kern-.1667em\lower.7ex\hbox{E}\kern-.125emX}}
    
\allowdisplaybreaks
    
\newcommand{\Rbb}{\mathbb{R}}

\mathchardef\myhyphen="2D

\newcommand{\xbarhat}{\hat{\xbar}}
\newcommand{\xShat}{\hat{\xvec}_S}
\newcommand{\xChat}{\hat{\xvec}_C}
\newcommand{\xFhat}{\hat{\xvec}_F}
\newcommand{\Qbar}{\bar{\Qmat}}
\newcommand{\qbar}{\bar{q}}
\newcommand{\WS}{\Wmat_{\!S}}
\newcommand{\WC}{\Wmat_{\!C}}
\newcommand{\WF}{\Wmat_{\!F}}
\newcommand{\Wbar}{\bar{\Wmat}}
\newcommand{\WbarS}{\Wbar_{\!S}}
\newcommand{\WbarC}{\Wbar_{\!C}}
\newcommand{\WbarF}{\Wbar_{\!F}}

\newcommand{\RF}{\Rmat_F}

\newcommand{\Jtilde}{J}

\newcommand{\Amat}{\matr{A}}
\newcommand{\Atilde}{\Tilde{\matr{A}}}
\newcommand{\Abar}{\bar{\matr{A}}}
\newcommand{\abar}{\bar{\matr{a}}}
\newcommand{\atilde}{\tilde{\matr{a}}}
\newcommand{\AS}{\matr{A}_{S}}
\newcommand{\AC}{\matr{A}_{C}}
\newcommand{\AF}{\matr{A}_{F}}

\newcommand{\Ka}{\Kmat_{\avec}}
\newcommand{\Kaf}{\Kmat_{\avec_F}}
\newcommand{\Kx}{\Kmat_{\xvec}}

\newcommand{\Kxbar}{\Kmat_{\xbar}}
\newcommand{\KxS}{\Kmat_{\xS}}
\newcommand{\KxC}{\Kmat_{\xC}}

\newcommand{\Kv}{\Kmat_{\vvec}}

\newcommand{\KxCxS}{\Kmat_{\xC\xS}}

\newcommand{\Khatxbar}{\Khat_{\xbar}}
\newcommand{\Khatvbar}{\Khat_{\vbar}}

\newcommand{\sigmav}{\sigma_{v}}

\newcommand{\sigmahatvbar}{\hat{\sigma}_{\bar{v}}}
\newcommand{\sigmahatvbaropt}{\hat{\sigma}_{\bar{v}*}}
\newcommand{\muM}{\bar{\mu}}
\newcommand{\muMtilde}{\Tilde{\bar{\mu}}}

\newcommand{\xhat}{\hat{\xvec}}

\newcommand{\xbar}{\bar{\xvec}}
\newcommand{\pbar}{\bar{p}}

\newcommand{\yhat}{\hat{\yvec}}

\newcommand{\egen}{\varepsilon}

\newcommand{\matr}[1]{\bm{#1}}

\newcommand{\Kmat}{\matr{K}}
\newcommand{\Lmat}{\matr{L}}
\newcommand{\Mmat}{\matr{M}}

\newcommand{\Qmat}{\matr{Q}}
\newcommand{\Rmat}{\matr{R}}
\newcommand{\Smat}{\matr{S}}
\newcommand{\Umat}{\matr{U}}
\newcommand{\Vmat}{\matr{V}}
\newcommand{\Wmat}{\matr{W}}

\newcommand{\Lambdamat}{\matr{\Lambda}}

\newcommand{\avec}{\matr{a}}

\newcommand{\vvec}{\matr{v}}
\newcommand{\vbar}{\bar{\matr{v}}}
\newcommand{\wvec}{\matr{w}}
\newcommand{\xvec}{\matr{x}}
\newcommand{\xS}{\matr{x}_S}
\newcommand{\xC}{\matr{x}_C}
\newcommand{\xF}{\matr{x}_F}
\newcommand{\yvec}{\matr{y}}
\newcommand{\zvec}{\matr{z}}

\newcommand{\Imat}{\matr{I}}

\newcommand{\eye}[1]{\Imat_{#1}}

\newcommand{\zerobf }{\boldsymbol 0}

\DeclareMathOperator{\Ebb}{\mathbb{E}}
\newcommand{\eunder}[1]{\underset{#1}{\Ebb}}

\DeclareMathAlphabet{\mymathbb}{U}{BOONDOX-ds}{m}{n}

\newcommand{\diag}{\text{diag}}

\newcommand{\T}{^\mathrm{T}}
\newcommand{\p}{^+}
\newcommand{\pT}{^{+\mathrm{T}}}

\newcommand{\inv}{^{-1}}

\let\oldin\in
\renewcommand{\in}{{\,\oldin\,}}

\let\oldnotin\notin
\renewcommand{\notin}{{\,\oldnotin\,}}

\renewcommand{\th}{\textsuperscript{th}}

\renewcommand{\tr}{\operatorname{tr}}

\long\def\/*#1*/{}

\newcommand{\Ic }{\mathcal{I}}

\newcommand{\Nc }{\mathcal{N}}

\definecolor{myblue}{rgb}{0.3328, 0.3539, 0.7758}
\definecolor{myblue2}{rgb}{0.0328, 0.0539, 0.4758}
\definecolor{mygreen2}{rgb}{ 0.0328 0.4758 0.0539} 
\definecolor{mygreen3}{rgb}{ 0.0328 0.1758 0.0539} 
\definecolor{myred}{rgb}{0.4758, 0.0328, 0.0539}
\definecolor{myred2}{rgb}{0.75, 0.0328, 0.0539}

\newcommand{\balert}[1]{{\color{myblue}{#1}}}

\renewcommand{\balert}[1]{{{#1}}}

\newcommand{\Khat}{\hat{\Kmat}}

\begin{document}
    
    \bstctlcite{IEEEexample:BSTcontrol}
    
    \title{\balert{Estimation under Model Misspecification with \\ Fake Features}}
    \author{
        \IEEEauthorblockN{
            Martin Hellkvist, 
            Ay\c ca \"Oz\c celikkale, 
            Anders Ahl\'{e}n
        }
        \IEEEauthorblockA{
            {Dept. of Electrical Engineering},
            {Uppsala University}, Sweden \\
            \{Martin.Hellkvist, 
            Ayca.Ozcelikkale, 
            Anders.Ahlen\}@angstrom.uu.se
        }
    }
    
    \maketitle
    
    \begin{abstract}
    We consider estimation under model misspecification
    where there is a model mismatch between the underlying system,
    which generates the data,
    and the model used during estimation.
   \balert{%
    We propose a model misspecification framework which enables a joint treatment of the model misspecification types of having fake features as well as incorrect covariance assumptions on the unknowns and the noise.
    We present a decomposition of the output error into components that relate to different subsets of the model parameters corresponding to underlying, fake and missing features.
    Here, 
    fake features are features which are included in the model but are not present in the underlying system.
    Under this framework,
    we characterize the estimation performance 
    and reveal trade-offs between the number of samples, 
    number of fake features,
    and the possibly incorrect noise level assumption.
    In contrast to existing work focusing on incorrect covariance assumptions or missing features, 
    fake features is a central component of our framework. 
    }%
    Our results show that fake features can significantly improve the estimation performance,
    even though they are not correlated with the features in the underlying system.
    In particular, 
    we show that the estimation error can be decreased by including more fake features in the model, 
    even to the point where the model is overparametrized, 
    i.e., the model contains more unknowns than observations.
\end{abstract}

    
    \newtheorem{thm}{\bf{Theorem}}
\newtheorem{cor}{\bf{Corollary}}
\newtheorem{lem}{\bf{Lemma}}
\newtheorem{prop}{\bf{Proposition}}
\newtheorem{rem}{\bf{Remark}}

\theoremstyle{remark} 
\newtheorem{defn}{\bf{Definition}}[section]
\newtheorem{ex}{\bf{Example}}[section]
\newtheorem{myexp}{\bf{Experiment}}

\newenvironment{theorem}
{\par\noindent \thm \begin{itshape}\noindent}
{\end{itshape} \vspace{3pt}}

\newenvironment{lemma}
{\par\noindent  \lem \begin{itshape}\noindent}
{\end{itshape} \vspace{3pt}}

\newenvironment{corollary}
{\par\noindent  \cor \begin{itshape}\noindent}
{\end{itshape} \vspace{3pt}
}

\newenvironment{remark}
{\par\noindent \rem \begin{itshape}\noindent}
{\end{itshape} \vspace{3pt}}

\newenvironment{definition}
{\par\noindent \defn \begin{itshape}\noindent}
{\end{itshape}}

\newenvironment{experiment}
{\vspace{3pt} \par\noindent \myexp \begin{itshape} \noindent}
{\end{itshape}\vspace{6pt}}

\newenvironment{example}
{\vspace{2pt} \par\noindent \ex} 
{\vspace{2pt}}

    \section{Introduction}\label{sec:introduction}
In this paper,
we study the linear estimation problem
when the model is misspecified, 
or in other words,
when there is a mismatch between between the true underlying system and the assumed model during estimation.  
In particular, 
the unknowns $\xS \in \Rbb^{p_S\times 1}$
and $\xC \in \Rbb^{p_C\times 1}$ generate the noisy observations $\yvec  \approx \AS \xS + \AC\xC $,
with the features in
$\AS\in\Rbb^{n\times p_S}$ and 
$\AC\in\Rbb^{n\times p_C}$. 
We consider the setting where this true underlying system is not fully known,
and $\AC$ is unavailable for estimation; hence $\AC$ constitute the missing features.
Moreover, 
the assumed model is misspecified with the fake features $\AF\in\Rbb^{n\times p_F}$. 
Hence, 
the mismatched estimator is based on the misspecified model $\yvec \approx \AS\xS + \AF\xF$.

\balert{%
    Within this framework,
    we analyze the effect that the fake features $\AF$ have on the estimation performance.}
    Here,
    the fake features $\AF$ are statistically independent of the underlying features $\AS$ and $\AC$,
    hence they have no explanatory power for the observations $\yvec$.
    This  formulation allows us to explore the fundamental relationships between the estimation performance and the model complexity.
    In particular, 
    we study
    the mismatch between the number of parameters in the true underlying system,
    and the number of parameters in the model assumed during estimation \cite{rao_misspecification_1971, breiman_how_nodate, belkin_reconciling_2019, belkin_two_2019,hellkvist_ozcelikkale_2021_modelmismatch_lmmse_IEEE,  hastie_surprises_2020,Bartlett_benign_overfitting_2020,muthukumar_harmless_2020,dascoli_more_2021,richards2021asymptotics,nakkiran2020optimal,mei_montanari_generalization_2021,lejeune_implicit_2020}. 
    
    Conventional wisdom suggests that  overparameterization,
    i.e., 
    when the number of tuneable parameters of a model exceeds the number of data points, 
    degrades the estimation performance. 
    On the other hand, recent work on double-descent curves has highlighted how the estimation performance
    can improve as the model complexity increases beyond the point of overparametrization
    \cite{belkin_reconciling_2019,belkin_two_2019,hellkvist_ozcelikkale_2021_modelmismatch_lmmse_IEEE,hastie_surprises_2020,muthukumar_harmless_2020,dascoli_more_2021,nakkiran2020optimal,Bartlett_benign_overfitting_2020,richards2021asymptotics,mei_montanari_generalization_2021}.
    \balert{%
    This phenomenon has been illustrated with various real-world datasets, 
    such as MNIST \cite[Fig.~2-4]{belkin_reconciling_2019},
    CIFAR-10 \cite[Fig.~S1]{belkin_reconciling_2019},
    TIMIT \cite[Fig.~S2]{belkin_reconciling_2019},
    and a wide range of estimators, 
    including neural networks \cite[Fig.~3]{belkin_reconciling_2019}, 
    random forests \cite[Fig.~4]{belkin_reconciling_2019}.
    }%
\balert{%
    In this paper, we contribute to the line of work on overparameterization by focusing on fake features and error decomposition under a statistical estimation framework. 
    }

Double-descent curves are often observed under some type of model misspecification,
such as missing features
\cite{breiman_how_nodate,belkin_two_2019,hastie_surprises_2020,hellkvist_ozcelikkale_2021_modelmismatch_lmmse_IEEE}
or incorrectly tuned regularization parameters \cite{nakkiran2020optimal}.
In \cite{rao_misspecification_1971},
a preliminary characterization of the least-squares estimation error with fake and missing features
has been presented 
for deterministic unknowns.
Under isotropic Gaussian features,
and with only missing features,
i.e., $p_F=0$,
more explicit expressions for the estimation error with deterministic priors are presented in \cite{breiman_how_nodate} and \cite{belkin_two_2019},
and in 
\cite{hellkvist_ozcelikkale_2021_modelmismatch_lmmse_IEEE}
with stochastic priors.
More general covariance structures have been investigated for the features \cite{hastie_surprises_2020,muthukumar_harmless_2020},
and connections between the estimation performance and the effective rank of the covariance matrix have been studied 
\cite{Bartlett_benign_overfitting_2020,AbuMostafa2012LearningFD}.
Misalignments between the features and unknowns have been studied for neural networks \cite{dascoli_more_2021}, 
as well as in the asymptotic regime of linear regression with a special case of noisy fake features \cite{richards2021asymptotics}. 
Mismatches between linear data generation and  non-linear estimation models have been investigated 
\cite{mei_montanari_generalization_2021}.
Double-descent curves have also been observed in the setting of distributed learning,
where each learner has a subset of the underlying features at its disposal, 
effectively creating local models with missing features and without fake features \cite{hellkvist_ozcelikkale_ahlen_linear_2021, HellkvistOzcelikkaleAhlen_distributed2020_spawc}.

Model misspecifications often arise in practical application scenarios
and can drastically affect the performance, such as in the case of positioning problems \cite{li_performance_2019},
channel estimation for wireless communications \cite{thanh_misspecified_2021}
and radar applications \cite{ren_performance_2015}.
Accordingly, 
robust estimation methods under covariance matrix uncertainties  have been  investigated  under a range of estimation settings,
such as  the constrained minimum mean squared error estimator \cite{lederman_tabrikian_constrained_2006},
the generalized difference regret criterion \cite{mittelman_miller_robust_2010}
and a maximum a-posteriori estimator  \cite{Zachariah_Shariati_Bengtsson_Estimation_2014},
as well as robustness under missing features \cite{Liu_Zachariah_Stoica_Robust_2020}.

An important example of the misspecification scenarios covered by our formulation,
is that of overcomplete dictionaries, 
i.e., dictionaries that contain more features than what is needed to represent a given family of signals.
In our formulation,
an overcomplete dictionary is obtained when there are fake features,
but no missing features.
Overcomplete dictionaries are commonly used for signal representation \cite{foucartRauhut_2013},
and such representations has been investigated from various aspects, 
including finding sparse representations within large overcomplete dictionaries in the compressive sensing framework \cite{foucartRauhut_2013},
and also under dictionary mismatch \cite{ChiScharfPezeshkiCalderbank_2011,TanYangNehorai_2014,BernhardtBoyerMarcosLarzabal_2016}. 
In contrast to the above line of work,
where a low-dimensional representation is sought, 
we instead focus on the scenario where the overcomplete dictionary is directly used.

The aim of this paper is to characterize 
the estimation performance with the model misspecifications of fake and missing features.
We formulate this problem in the framework of linear minimum mean squared error (LMMSE) estimation \cite{kailath_sayed_hassibi_linear_estimation_2000}.
We focus on the average estimation performance in terms of the unknowns, 
i.e., the mean squared error (MSE).
Another type of model misspecification that we consider is constituted by potential discrepancies between the true covariance matrices and those used during estimation.
The main contributions of this work are as follows:
\begin{itemize}
    \item \balert{We present a statistical estimation framework which introduces a general notion of model misspecification
    with both fake features
    and
    incorrect covariance matrix assumptions.}
    
    \item \balert{ We provide  analytical expressions for the
    MSE related to each set of unknowns $\xS$, $\xC$ and $\xF$ as well as the output $\yvec$,
    as functions of the number of fake features and missing features,
    for both under- and overparameterized settings,
    i.e., $n > p_S + p_F $ and $n < p_S + p_F$.}

    \item \balert{We show that the presence of fake features can significantly improve the estimation performance for $\xS$
    compared to when there are no fake features; 
    even in the overparametrized regime,
    and even though the fake features are uncorrelated with the true underlying features. }%
    
    \item \balert{Our decomposition of the MSE associated with $\yvec$ in terms of the MSEs associated with $\xS$, $\xC$ and $\xF$  quantifies the difference in the behaviour of these errors. }

\end{itemize}

\balert{Consistent with the line of work focusing on missing features \cite{breiman_how_nodate,belkin_two_2019,hellkvist_ozcelikkale_2021_modelmismatch_lmmse_IEEE, hastie_surprises_2020},
we show that the error can be very large if the number of observations $n$ is close to the assumed model dimension,}
which in our setting is $p_S + p_F$,
i.e., the sum of the number of included underlying features and fake features.
The decrease in error with increasing $p_F$ in the overparametrized regime confirms  the presence of double-descent behaviour even with fake features.
\balert{%
     Our results further show that even when the MSE associated with $\yvec$ stays on the same level for the under/over-parameterized scenarios,
     the minimum error for the unknowns $\xS$ can be obtained in the overparametrized scenario with a high number of fake features. 
     These results also suggest that the model may learn the parameters of the underlying system even though the output prediction is poor. 
}

The rest of the paper is structured as follows:
Section~\ref{sec:problem} presents the problem formulation. 
In Section~\ref{sec:analytical},
we provide our first main result in Theorem~\ref{thm:emse_noisfree},  which characterizes the MSE when the assumed noise level is zero.
The problem setting is then generalized in Section~\ref{sec:analytical_II} with a non-zero noise level assumption,
and we present our second main result in Theorem~\ref{thm:emse_noise_model}.
We illustrate our results with numerical experiments in Section~\ref{sec:numerical}. We provide further discussions in Section~\ref{sec:discussions},
and conclude the paper in Section~\ref{sec:conclusions}.

\textbf{Notation:}
        We denote the Moore-Penrose pseudoinverse and the transpose of a matrix $\Amat$ as $\Amat\p$ and $\Amat\T$, respectively.
        The $m\times m$ identity matrix is denoted as $\eye{m}$.
        The Euclidean norm and trace operator are denoted by $\|\cdot\|$ and  $\tr(\cdot)$, respectively.
        We use $\eunder{\xvec}[\cdot]$ or $\Ebb_{\xvec}[\cdot]$ to emphasize that the expectation is taken with respect to the random variable $\xvec$.
        For two random column vectors
        $\zvec$, $\wvec$, we denote their covariance matrix by $\Kmat_{\zvec\wvec} = \Ebb_{\zvec,\wvec}[(\zvec - \Ebb_{\zvec}[\zvec]) (\wvec-\Ebb_{\wvec}[\wvec])\T]$.
        For auto-covariance matrices, 
        we write the subscript only once: $\Kmat_{\zvec} = \Kmat_{\zvec\zvec}$. 
        We use the notation $(\cdot)_+=\max(\cdot,0)$.
        We refer a vector/matrix with  independent and identically distributed (i.i.d.) elements with $\Nc(0,1)$ as a standard Gaussian random vector/matrix. 

    \section{Problem Statement}
\label{sec:problem}

\subsection{The Underlying System and the Misspecified Model}
\label{sec:problem:model_and_misspec}

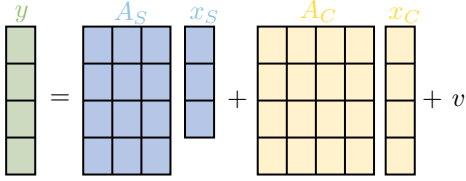
\begin{figure}[t]
    \centering
    \tikzset{every picture/.style={line width=0.75pt}} 

\begin{tikzpicture}[x=0.55pt,y=0.7pt,yscale=-1,xscale=1] 

\draw  [draw opacity=0][fill={rgb, 255:red, 205; green, 218; blue, 191 }  ,fill opacity=1 ] (38,191) -- (58,191) -- (58,271) -- (38,271) -- cycle ; \draw    ; \draw   (38,211) -- (58,211)(38,231) -- (58,231)(38,251) -- (58,251) ; \draw   (38,191) -- (58,191) -- (58,271) -- (38,271) -- cycle ;
\draw  [draw opacity=0][fill={rgb, 255:red, 182; green, 198; blue, 231 }  ,fill opacity=1 ] (91,191) -- (151,191) -- (151,271) -- (91,271) -- cycle ; \draw   (111,191) -- (111,271)(131,191) -- (131,271) ; \draw   (91,211) -- (151,211)(91,231) -- (151,231)(91,251) -- (151,251) ; \draw   (91,191) -- (151,191) -- (151,271) -- (91,271) -- cycle ;
\draw  [draw opacity=0][fill={rgb, 255:red, 182; green, 198; blue, 231 }  ,fill opacity=1 ] (161,191) -- (181,191) -- (181,251) -- (161,251) -- cycle ; \draw    ; \draw   (161,211) -- (181,211)(161,231) -- (181,231) ; \draw   (161,191) -- (181,191) -- (181,251) -- (161,251) -- cycle ;
\draw  [draw opacity=0][fill={rgb, 255:red, 255; green, 242; blue, 205 }  ,fill opacity=1 ] (211,191) -- (291,191) -- (291,271) -- (211,271) -- cycle ; \draw   (231,191) -- (231,271)(251,191) -- (251,271)(271,191) -- (271,271) ; \draw   (211,211) -- (291,211)(211,231) -- (291,231)(211,251) -- (291,251) ; \draw   (211,191) -- (291,191) -- (291,271) -- (211,271) -- cycle ;
\draw  [draw opacity=0][fill={rgb, 255:red, 255; green, 242; blue, 205 }  ,fill opacity=1 ] (299,191) -- (319,191) -- (319,271) -- (299,271) -- cycle ; \draw    ; \draw   (299,211) -- (319,211)(299,231) -- (319,231)(299,251) -- (319,251) ; \draw   (299,191) -- (319,191) -- (319,271) -- (299,271) -- cycle ;

\draw (66,226) node [anchor=north west][inner sep=0.75pt]    {$=$};
\draw (188,224) node [anchor=north west][inner sep=0.75pt]    {$+$};
\draw (110,176) node [anchor=north west][inner sep=0.75pt]  [color={rgb, 255:red, 144; green, 195; blue, 225 }  ,opacity=1 ]  {$A_{S}$};
\draw (238,175) node [anchor=north west][inner sep=0.75pt]  [color={rgb, 255:red, 255; green, 220; blue, 62 }  ,opacity=1 ]  {$A_{C}$};
\draw (299,179) node [anchor=north west][inner sep=0.75pt]  [color={rgb, 255:red, 255; green, 220; blue, 62 }  ,opacity=1 ]  {$x_{C}$};
\draw (162,179) node [anchor=north west][inner sep=0.75pt]  [color={rgb, 255:red, 144; green, 195; blue, 225 }  ,opacity=1 ]  {$x_{S}$};
\draw (42,178) node [anchor=north west][inner sep=0.75pt]  [color={rgb, 255:red, 151; green, 190; blue, 110 }  ,opacity=1 ]  {$y$};
\draw (322,224) node [anchor=north west][inner sep=0.75pt]    {$+\ v$};

\end{tikzpicture}
    \caption{ 
        Illustration of the underlying system. 
        The features in $\AS$ and $\AC$  together with the noise $\vvec$ generate the observations $\yvec$.
        }
    \kern-0.4em
    \label{fig:sysfig_true}
\end{figure}

\begin{figure}[t]
    \centering
    \tikzset{every picture/.style={line width=0.75pt}} 

\begin{tikzpicture}[x=0.55pt,y=0.7pt,yscale=-1,xscale=1] 

\draw  [draw opacity=0][fill={rgb, 255:red, 205; green, 218; blue, 191 }  ,fill opacity=1 ] (58,194) -- (78,194) -- (78,274) -- (58,274) -- cycle ; \draw    ; \draw   (58,214) -- (78,214)(58,234) -- (78,234)(58,254) -- (78,254) ; \draw   (58,194) -- (78,194) -- (78,274) -- (58,274) -- cycle ;
\draw  [draw opacity=0][fill={rgb, 255:red, 182; green, 198; blue, 231 }  ,fill opacity=1 ] (111,194) -- (171,194) -- (171,274) -- (111,274) -- cycle ; \draw   (131,194) -- (131,274)(151,194) -- (151,274) ; \draw   (111,214) -- (171,214)(111,234) -- (171,234)(111,254) -- (171,254) ; \draw   (111,194) -- (171,194) -- (171,274) -- (111,274) -- cycle ;
\draw  [draw opacity=0][fill={rgb, 255:red, 182; green, 198; blue, 231 }  ,fill opacity=1 ] (181,194) -- (201,194) -- (201,254) -- (181,254) -- cycle ; \draw    ; \draw   (181,214) -- (201,214)(181,234) -- (201,234) ; \draw   (181,194) -- (201,194) -- (201,254) -- (181,254) -- cycle ;
\draw  [draw opacity=0][fill={rgb, 255:red, 255 ; green, 230; blue, 225 }  ,fill opacity=1 ] (231,194) -- (311,194) -- (311,274) -- (231,274) -- cycle ; \draw   (251,194) -- (251,274)(271,194) -- (271,274)(291,194) -- (291,274) ; \draw   (231,214) -- (311,214)(231,234) -- (311,234)(231,254) -- (311,254) ; \draw   (231,194) -- (311,194) -- (311,274) -- (231,274) -- cycle ;
\draw  [draw opacity=0][fill={rgb, 255:red, 255; green, 230; blue, 225 }  ,fill opacity=1 ] (319,194) -- (339,194) -- (339,274) -- (319,274) -- cycle ; \draw    ; \draw   (319,214) -- (339,214)(319,234) -- (339,234)(319,254) -- (339,254) ; \draw   (319,194) -- (339,194) -- (339,274) -- (319,274) -- cycle ;

\draw (87,229) node [anchor=north west][inner sep=0.75pt]    {$=$};
\draw (208,227) node [anchor=north west][inner sep=0.75pt]    {$+$};
\draw (342,227) node [anchor=north west][inner sep=0.75pt] {$+\ \overline{v}$};
\draw (130,178) node [anchor=north west][inner sep=0.75pt]  [color={rgb, 255:red, 144; green, 195; blue, 225 }  ,opacity=1 ]  {$A_{S}$};
\draw (258,178) node [anchor=north west][inner sep=0.75pt]  [color={rgb, 255:red, 225; green, 125; blue, 125 }  ,opacity=1 ]  {$A_{F}$};
\draw (319,181) node [anchor=north west][inner sep=0.75pt]  [color={rgb, 255:red, 225; green, 125; blue, 125 }  ,opacity=1 ]  {$x_{F}$};
\draw (182,181) node [anchor=north west][inner sep=0.75pt]  [color={rgb, 255:red, 144; green, 195; blue, 225 }  ,opacity=1 ]  {$x_{S}$};
\draw (62,180) node [anchor=north west][inner sep=0.75pt]  [color={rgb, 255:red, 151; green, 190; blue, 110 }  ,opacity=1 ]  {$y$};

\end{tikzpicture}
    \caption{ 
        Illustration of the misspecified model. 
        The features in $\AC$ are missing,
        and there are additional fake features $\AF$.
        }
    \kern-0.4em
    \label{fig:sysfig_false}
\end{figure}
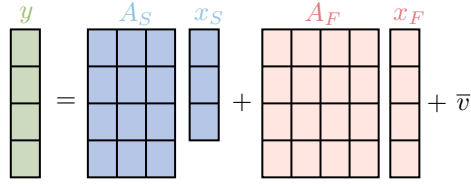

    In this paper, 
    we consider noisy observations $\yvec$ which are generated by the following linear system,
    \begin{equation}\label{eqn:model_true}
        \yvec = \Atilde\xvec + \vvec = \AS\xS + \AC\xC + \vvec.
    \end{equation}
    We refer to the system in \eqref{eqn:model_true} as the 
    \textit{underlying system}.
    The following linear model represents a misspecification of the underlying system 
    \begin{equation}\label{eqn:model_false}
        \yvec = \Abar\xbar + \vbar = \AS\xS + \AF\xF + \vbar. 
    \end{equation}
    We refer to \eqref{eqn:model_false} as the \textit{misspecified model}.
    Here, the matrix
        $\AS \in \Rbb^{n \times p_S}$ represents the features that are present in both \eqref{eqn:model_true} and \eqref{eqn:model_false}, and the matrices
        $\AC \in \Rbb^{n \times p_C}$ and
        $\AF \in \Rbb^{n \times p_F}$
        represent
        the missing features
        and the fake features,
        respectively.
        The matrices $\Atilde$ and $\Abar$ are composed as
        \begin{equation}
        \Atilde = [\AS, \, \AC] \in \Rbb^{n \times p},
        \end{equation}
        \begin{equation}
        \Abar = [\AS, \, \AF] \in \Rbb^{n \times \pbar}.
        \end{equation}
        The unknowns are denoted by
        $\xS \in \Rbb^{p_S \times 1}$,
        $\xC \in \Rbb^{p_C \times 1}$, and
        $\xF \in \Rbb^{p_F \times 1}$, respectively,
        where
        \begin{equation}\label{eq:def:xvec}
            \xvec = [\xS\T, \, \xC\T]\T \in \Rbb^{p \times 1},
        \end{equation}
        \begin{equation}
            \xbar = [\xS\T, \, \xF\T]\T \in \Rbb^{\pbar \times 1}.
        \end{equation}
    We have that $p = p_S + p_C$ and $\pbar = p_S + p_F$.
    The vector $\yvec\in\Rbb^{n\times1}$ denotes the observations,
    and $\vvec\in\Rbb^{n\times 1}$ and $\vbar \in \Rbb^{n\times 1}$ denote the observation noise in the underlying system and the misspecified model,
    respectively.
    Figure~\ref{fig:sysfig_true} and Figure~\ref{fig:sysfig_false} illustrate the underlying system and the misspecified model, respectively. 
    Note that  the underlying system in  \eqref{eqn:model_true} can be equivalently expressed as follows, 
    by setting $\xF\triangleq \bm{0}$,
     \begin{equation}
        \yvec = \AS\xS + \AC\xC + \AF\xF + \vvec,
    \end{equation}
    highlighting the mismatch between 
     the underlying system and the misspecified model \eqref{eqn:model_false}. 
    
    In the considered setting,
    the observations in $\yvec$ are generated by the underlying system in \eqref{eqn:model_true} and we are interested in estimating the unknowns $\xvec$. 
    We do not have full knowledge of the underlying system in \eqref{eqn:model_true}.
    Instead, we base the estimation on the belief that $\yvec$ comes from the 
    (incorrect, misspecified) 
    model in \eqref{eqn:model_false}.
    In particular,
    we do not have full knowledge of the correct feature matrix $\Atilde=[\AS,\AC]$.
    In the misspecified model, not only the features in $\AC$ are missing, but there are also additional fake features $\AF$ that are not present in the underlying system.

    The unknowns $\xS$, $\xC$, and $\xF$,
    and the noise vectors $\vvec$ and $\vbar$ are zero-mean random vectors. 
    The noise vectors are uncorrelated with each other and the unknowns. 
    For any random vector $\zvec \in \Rbb^m$, the associated covariance matrix is denoted by $\Kmat_{\zvec} \in \Rbb^{m \times m}$.   

    In addition to using the
    misspecified model in \eqref{eqn:model_false},
    the true covariance matrices are not known during estimation,
    and we use $\Khat$ to denote an assumed covariance matrix.
    For example, $\Khatxbar$,
    denotes the assumed covariance matrix of $\xbar$.
    Note that the assumed covariance matrix is not necessarily equal to its true counterpart $\Kxbar$,
    i.e., we possibly have $\Khatxbar \neq \Kxbar$.


\subsection{Linear Estimation and the MSE}
\label{sec:prob_stat:linear_est_mse}
    
    We consider linear estimators,
    i.e.,
    estimators such that the estimate 
    is a linear function of the observations. 
    Suppose that $(\WS,\WC,\WF)$ are estimators for
    $(\xS,\xC,\xF)$,
    with the associated estimates 
    $\xShat = \WS \yvec$, 
    $\xChat = \WC \yvec$ and
    $\xFhat = \WF \yvec$.  
    Then, the corresponding MSEs
    are given by
    \begin{align}\label{eqn:JS_WS}
        J_S(\WS)
        &= \eunder{\xvec, \yvec}\!\left[\|\xS - \xShat\|^2\right]
        = \eunder{\xvec, \yvec}\!\left[\|\xS - \WS\yvec\|^2\right]\!,\\
        J_C(\WC)
        &= \eunder{\xvec, \yvec}\!\left[\|\xC - \xChat\|^2\right]
        = \eunder{\xvec, \yvec}\!\left[\|\xC - \WC\yvec\|^2\right]\!,\label{eqn:JC_WC}\\
        J_F(\WF)
        &= \eunder{\xvec, \yvec}\!\left[\|\xF - \xFhat\|^2\right]
        = \eunder{\xvec, \yvec}\!\left[\|\xF - \WF\yvec\|^2\right]\!,\label{eqn:JF_WF}
    \end{align}
    where the expectations are taken over the probability distributions of $\xvec$ and $\yvec$ from the underlying system in \eqref{eqn:model_true}. 
    The estimators for $\xvec = [\xS\T,\xC\T]\T$ and $\xbar =[\xS\T, \, \xF\T]\T$
    can be written as $\Wmat = [\WS\T,\,\WC\T]\T \in \Rbb^{p\times n}$ and 
    $\Wbar = [\WbarS\T,\,\WbarF\T]\T \in \Rbb^{\pbar\times n}$,
    with the associated estimates  $\xhat = \Wmat \yvec$ 
    and $\xbarhat = \Wbar \yvec$.
    Hence, the estimators for the composite vectors $\xvec$ and $\xbar$ are written as a combination of the linear estimators of their respective parts. 

    Our interest lies in the unknowns of the underlying system $\xvec = [\xS\T,\xC\T]\T$.
    Hence we focus on the MSE over $\xvec$, i.e.,
    \begin{equation}\label{eqn:mse}
        \Jtilde(\Wmat) 
        = \eunder{\xvec,\yvec}\left[\|\xvec - \xhat\|^2\right]
        = J_S(\Wmat_S) + J_C(\Wmat_C).
    \end{equation}

\subsection{LMMSE Estimation -- True Model}\label{sec:prob_stat:lmmse_true}

    The LMMSE estimator, 
    i.e., the matrix $\Wmat\in\Rbb^{p\times n}$ 
    that minimizes the MSE $J(\Wmat)$, 
    is given by 
    $\Wmat_O = \arg\min_{\Wmat}\Ebb_{\xvec, \yvec}\!\left[\|\xvec - \Wmat \yvec\|^2\right]$, 
    where the subscript $O$ emphasizes that the estimator is optimal. 
    Under correct model assumptions,
    $\Wmat_O$
    is given by,
    \cite{kailath_sayed_hassibi_linear_estimation_2000},
    \begin{align}\label{eqn:lmmse_true}
        \!\!
        \Wmat_O \! &= \! \Kmat_{\xvec \yvec} \Kmat_{\yvec}\p 
        \! = \! \Kmat_{\xvec} \Atilde\T 
        ( \Atilde \Kmat_{\xvec} \Atilde\T \!\!
        + \! \Kmat_{\vvec} )\p
        \!\! = \!\! \begin{bmatrix}
            \Wmat_{O,S} \\
            \Wmat_{O,C}
        \end{bmatrix}.
    \end{align}
    Here we have used that $\yvec$ comes from the underlying system in \eqref{eqn:model_true}, 
    from which
    $\Kmat_{\xvec \yvec} = \Kmat_{\xvec} \Atilde\T\in\Rbb^{p\times n}$,
    and $\Kmat_{\yvec} = \Atilde \Kmat_{\xvec} \Atilde\T + \Kmat_{\vvec}\in\Rbb^{n\times n}$.
    The matrices $\Wmat_{O,S}\in\Rbb^{p_S\times n}$ and 
    $\Wmat_{O,C}\in\Rbb^{p_C\times n}$ denote the blocks of $\Wmat_O$ which estimate $\xS$ and $\xC$, respectively.
    
    In \eqref{eqn:lmmse_true},
    we use the Moore-Penrose pseudoinverse, hence the estimator
    minimizes the MSE regardless of whether $\Kmat_{\yvec}$ is singular or not
    \cite[Theorem 3.2.3]{kailath_sayed_hassibi_linear_estimation_2000}.
    
\subsection{LMMSE Estimation -- Misspecified Model}\label{sec:prob_stat:lmmse_false}
    
    In this paper, 
    our focus 
    is 
    on estimation under the model mismatch caused by the discrepancy between 
    the underlying system \eqref{eqn:model_true} and the misspecified model \eqref{eqn:model_false}.
    Hence, we consider the estimator based on the assumed covariance matrices 
    $\Khat_{\xbar \yvec}\in\Rbb^{\pbar\times n}$ 
    and $\Khat_{\yvec}\in\Rbb^{n\times n}$,
    which are given by
    $\Khat_{\xbar \yvec} = \Khat_{\xbar} \Abar\T $ 
    and 
    $\Khat_{\yvec}=\Abar \Khat_{\xbar} \Abar\T + \Khat_{\vbar}$.
    The associated estimator $\Wbar\in\Rbb^{\pbar\times n}$
    is thus
    \begin{equation}\label{eqn:lmmse_false}
        \Wbar = \Khat_{\xbar \yvec} \Khat_{\yvec}\p
        = \Khat_{\xbar} \Abar\T 
        (\Abar \Khat_{\xbar} \Abar\T + \Khat_{\vbar})\p
        \! = \!\begin{bmatrix}
        \WbarS \\
        \WbarF
        \end{bmatrix},
    \end{equation}
    where $\WbarS\in\Rbb^{p_S\times n}$ and $\WbarF\in\Rbb^{p_F\times n}$.
    Given $\yvec$ from \eqref{eqn:model_true},
    the estimator $\Wbar$ produces the estimate
    \begin{equation}\label{eqn:xbarhat}
        \xbarhat 
        = \begin{bmatrix}
            \xShat \\ \xFhat
        \end{bmatrix}
        = \Wbar \yvec,
    \end{equation}
    where 
    $\xShat = \WbarS\yvec\in\Rbb^{p_S\times 1}$,    $\xFhat=\WbarF\yvec\in\Rbb^{p_F\times 1}$.
    Note that $\xC$ is missing from the misspecified model \eqref{eqn:model_false},
    hence the associated estimator of $\xC$ is set to 
    $\WbarC=\bm{0} \in\Rbb^{p_C\times n}$ and 
    we have $\xChat = \Ebb_{\xvec}[\xC]= \bm{0}$.
    The corresponding MSEs in 
    {\eqref{eqn:JS_WS} -- \eqref{eqn:JF_WF}} are then
    $J_S(\Wbar_S)$, 
    $J_C(\Wbar_C)$ and 
    $J_F(\Wbar_F)$.

    A key performance criterion considered in this paper is the estimation error for $\xS$ and $\xC$ of the underlying system.
    The corresponding MSE is given by
    \begin{equation}\label{eqn:mse_WbarC_zero}
        \Jtilde(\Wbar) 
        = J_S(\WbarS) + J_C(\WbarC).
    \end{equation}
    Note that the expectation in $\Jtilde (\Wbar)$ is taken with respect to the true underlying distribution of $\xvec$ and $\yvec$,
    hence it evaluates the MSE that is obtained when the data comes from \eqref{eqn:model_true} but the estimator $\Wbar$ is used, which is based on the misspecified model in \eqref{eqn:model_false}. 
    
    To summarize our setting,
    $\yvec$ is generated by the system in \eqref{eqn:model_true}.
    The estimation is performed using the features available to us, $\AS$ and $\AF$,
    and produce estimates for $\xS$ and $\xF$ based on the misspecified model in \eqref{eqn:model_false},
    implicitly setting the estimate of $\xC$ to zero.
    %
    
\subsection{Random Feature Matrices and Expected MSE}\label{sec:mse_random_A}
    \label{sec:problem_expected_MSE}
     We now define the expected MSE over features, which is the main performance criterion in this work.

    We analyze the MSE $\Jtilde(\Wbar)$ in \eqref{eqn:mse_WbarC_zero},
    and specifically how it depends on the feature matrices $\AS$, $\AC$ and $\AF$.
    We model the feature matrices as statistically independent standard Gaussian random matrices,
    i.e., all elements of the matrices are i.i.d. with $\Nc(0,1)$.
    The features are uncorrelated with the unknowns and the noise. 
    
    The \textit{expected MSE} 
    associated with $\Wbar$ in \eqref{eqn:lmmse_false} over the distribution of 
    $\Amat=[\AS,\AC,\AF] \in \Rbb^{n\times (p_S+p_C+p_F)}$
    is
    \begin{equation}\label{eqn:emse_false}
        \egen(p_S, p_C, p_F, n) 
        = \eunder{\Amat} \left[ \Jtilde(\Wbar) \right].
    \end{equation}  
    In other words,
    we obtain one set of feature matrices $[\AS,\AC,\AF]$,
    compute $\Wbar$,
    and the associated MSE $\Jtilde (\Wbar)$ with respect to the distribution of $\xvec$ and $\vvec$.
    We then compute the expected MSE $\egen$ as the expectation of $J(\Wbar)$
    with respect to the distribution of the features in $\Amat$.
    Hence, the statistical model on the features lets us analyze the MSE from the perspective of performing multiple experiments over different realizations of $\Amat$.

    \section{Expected MSE under Model Misspecification}
\label{sec:analytical}
In this section, we present one of our main results,
 Theorem~\ref{thm:emse_noisfree},
which provides an analytical expression for the expected MSE of the LMMSE estimator $\Wbar$ in \eqref{eqn:lmmse_false} under the assumption $\vbar=\zerobf$, 
i.e., the assumed noise level is zero: 

\begin{theorem}\label{thm:emse_noisfree}
    Let $\Khat_{\xbar} = \eye{\pbar}$, $\Kv=\sigmav^2\eye{n}$, $\Khat_{\vbar} = \zerobf$, $n\geq 1$.
    If $n > \pbar + 1 $,
    then the expected MSE in \eqref{eqn:emse_false} associated with $\Wbar$ in \eqref{eqn:lmmse_false} is
    \begin{equation}\label{eqn:thm:emse_noisefree:n_geq_pbar}
        \egen(p_S, p_C, p_F, n) =
        \tfrac{p_S}{n-\pbar-1}(\tr(\Kmat_{\xC}) + \sigmav^2)
        + \tr(\Kmat_{\xC}). 
    \end{equation}
    If $\pbar > n+1$, then
    \begin{align}\label{eqn:thm:emse_noisefree:n_less_pbar}
    \begin{split}
        &\egen(p_S, p_C, p_F, n) = 
        \tfrac{n p_S}{\pbar(\pbar - n -1)}
        (\tr(\Kmat_{\xC}) + \sigmav^2) \\
        &~~\qquad + \left(1 - \tfrac{n}{\pbar} 
        - \tfrac{p_F n (\pbar-n)}{(\pbar-1)\pbar(\pbar+2)}\right)\tr(\Kmat_{\xS})
        + \tr(\Kmat_{\xC}).
    \end{split}
    \end{align}
    
\end{theorem}

\noindent Proof: See Appendix~\ref{proof:thm:emse_noisfree}.

Recall that $\pbar = p_S + p_F$.
Theorem~\ref{thm:emse_noisfree} shows how the expected MSE $\egen$ varies with
the number of samples $n$,
observed underlying features $p_S$
and fake features $p_F$ in the misspecified model \eqref{eqn:model_false}.
In particular, Theorem~\ref{thm:emse_noisfree}  shows the following: 
\begin{enumerate}
    \item The expected MSE $\egen$ has no dependence on the covariance structure of the unknowns,
    but only depends on the respective power levels,
    i.e., $\tr(\KxS)$ and $\tr(\KxC)$.
    
    \item The expected MSE takes on very large values when the number of data points is close to the assumed model size, i.e., $n \approx \pbar$. See that $n-\pbar-1$ or $\pbar-n-1$ appear in the denominators of the respective leading terms in \eqref{eqn:thm:emse_noisefree:n_geq_pbar} and \eqref{eqn:thm:emse_noisefree:n_less_pbar}. 
    Note that this peak around $n \approx \pbar$ occurs only if  $\tr(\KxC) + \sigmav^2 > 0$, 
    i.e.,
    the observations are noisy, 
    or there are missing features.
    
    \item The expected MSE is not a monotonically increasing function of $p_F$. We further discuss this point in  Section~\ref{sec:thm1:effectofpf}.
    
    \item In the limit of $n\rightarrow\infty$,
    the effect of the fake features vanishes,
    and the observed unknowns $\xS$ are estimated perfectly.
    Hence the error approaches $\egen\rightarrow\tr(\KxC)$,
    i.e., the missing unknowns $\xC$ constitute all of the error.
    
\end{enumerate}

    \balert{%
    \begin{remark}\label{remark:interpolation_bad_estimate}
        Let $\Khatvbar=\bm 0$, $\Khatxbar \succ 0$. Consider a given vector of observations $\yvec$.
        If the model has as many parameters as there are observations,
        i.e., $\pbar = n$, 
        then the model perfectly fits the observations at hand:
        \begin{equation}
            \yhat = \Abar \xbarhat = \Abar \Wbar \yvec 
            = \Abar \Khatxbar \Abar\T (\Abar\Khatxbar\Abar\T)\inv \yvec
            = \yvec.
        \end{equation}
       On the other hand, $\xbarhat$ does not necessarily provide a good estimate of the unknowns; hence the large error values around $n \approx \pbar$ in \eqref{eqn:thm:emse_noisefree:n_geq_pbar} and \eqref{eqn:thm:emse_noisefree:n_less_pbar} are obtained.
    \end{remark} }
    
    The fake features in our framework can be associated with the weak features investigated in \cite{dascoli_more_2021, richards2021asymptotics}.   
    Consistent with \cite{dascoli_more_2021, richards2021asymptotics}, 
    our work shows that irrelevant features can be beneficial.

    \balert{%
    The main contribution of Theorem~\ref{thm:emse_noisfree} is  quantification of the fake features' effect on the estimation performance for the underlying features $\xvec_s$.
    }
    \balert{Theorem~\ref{thm:emse_noisfree} with $p_F>0$ together with the results on the output error 
    (i.e., the error in $y$, see \eqref{eqn:egen_y_decomposed}) 
    shows that the error $\egen$, 
    i.e.,
    the error for the model parameters $\xS$ and $\xC$,
    behaves significantly different from the output error. 
    This aspect 
    has been overlooked in the literature which either focuses on the output error,
    which is not necessarily the same as the error in the model parameters $\xS$ and $\xC$,
    or consider the unknowns without fake features.
    For the special case of $p_F=0$,   Theorem~\ref{thm:emse_noisfree} provides the output error (minus an additive term of $\sigmav^2$) which is consistent
    with the results with missing features   \cite{belkin_two_2019, hellkvist_ozcelikkale_2021_modelmismatch_lmmse_IEEE}. Nevertheless,  the results of  Theorem~\ref{thm:emse_noisfree} in its general form cannot be derived from expressions of the output error and
    it constitutes a component of a non-trivial decomposition of the output error, see Section~\ref{sec:LabelMSE}.
    }

\subsection{Effect of Fake Features}
\label{sec:thm1:effectofpf}

Theorem~\ref{thm:emse_noisfree} shows that the presence of fake features, i.e., $p_F>0$, can be beneficial to the estimation performance when the model is misspecified.
In this section,
we discuss this phenomenon. 

For values of $p_F$ such that $\pbar < n - 1$, 
i.e., if the misspecified model is underparameterized,
then $p_F=0$ minimizes the expected MSE $\egen$.
In other words,
$\egen$ is monotonically increasing with $p_F$ in the underparameterized regime.
The overparametrized case,
i.e., $\pbar > n + 1$, 
is less straightforward and discussed next.

\begin{corollary}\label{lemma:local_minima_overp}
    Consider the setting of Theorem~\ref{thm:emse_noisfree}.
    If the number of samples $n$, and underlying unknowns $p_S$ and $p_C$ are fixed,
    and $n<\infty$, $p_S<\infty$ and $p_C<\infty$, 
    then the following holds:
    \begin{enumerate}
        \item[i)]  
        $\lim_{p_F\rightarrow\infty} \egen(p_S,p_C,p_F,n) 
        = \tr(\Kx)$.
        
        \item[ii)] 
        If $p_F\rightarrow\infty$,
        then the expected MSE $\egen$ approaches the limit in i) from below.
    \end{enumerate}
    
\end{corollary}

\noindent Proof: See Appendix~\ref{proof:lemma:local_minima_overp}.

By Theorem~\ref{thm:emse_noisfree},
if $\tr(\KxC) + \sigmav^2 > 0$,
the expected MSE $\egen$ diverges if 
$n\approx \pbar$.
On the other hand, 
Corollary~\ref{lemma:local_minima_overp} shows that $\egen$ approaches its limit for ${p_F\rightarrow\infty}$ from below with increasing $p_F$. 
These observations together reveal the following: 
\begin{remark}
    Under $\tr(\KxC) + \sigmav^2 > 0$ and $\Khat_{\vbar} = \zerobf$, $\egen$ is non-monotonic as $p_F$ increases
    and there is a local minimum with non-zero $p_F$   in the overparametrized regime, i.e.,  $n<\pbar$.

\end{remark}

    The following corollary shows that under certain signal power conditions,
    the expected MSE $\egen$ is lower for $p_F\rightarrow\infty$
    than for $p_F=0$.

\begin{corollary}\label{lemma:power_ratio_limits}
    Consider the setting of Theorem~\ref{thm:emse_noisfree}.
    Let $\tr(\KxS)=r\tr(\Kx)$, and $\tr(\KxC)=(1-r)\tr(\Kx)$
    with $0\leq r \leq 1$.
    If $n\geq p_S$, $n>1$, and
    \begin{equation}\label{eqn:r_bound_n_geq_ps}
        r < \frac{p_S}{n-1}\frac{\tr(\Kx) + \sigmav^2}{\tr(\Kx)},
    \end{equation}
    or, if $n < p_S$, and
    \begin{equation}\label{eqn:r_bound_n_leq_ps}
        r < \frac{p_S}{2p_S -n -1}\frac{\tr(\Kx) + \sigmav^2}{\tr(\Kx)},
    \end{equation}
     then the expected MSE is smaller as $p_F\rightarrow\infty$ than for $p_F=0$.

\end{corollary}

\noindent Proof: See Appendix~\ref{proof:lemma:power_ratio_limits}.

With the insights gained from Corollary~\ref{lemma:local_minima_overp} and Corollary~\ref{lemma:power_ratio_limits}, 
we observe the following: 
\begin{remark}
    Even though the fake features represents a model misspecification,
    their presence can improve the estimation performance,
    even when the model with fake features is drastically overparameterized.
\end{remark}

In the setting of Theorem~\ref{thm:emse_noisfree},
the estimator $\Wbar$ from \eqref{eqn:lmmse_false} is given by $\Wbar = \Abar\p = (\Abar\T\Abar)\p\Abar\T$,
where $\Abar=[\AS,\AF]$ is the matrix of regressors of the misspecified model.
The underlying mechanism which explains the potential benefits of fake features is directly connected to the spectral properties of the matrix $\Abar\T\Abar$.
This point is discussed in more detail in the subsequent sections.

\section{Model Misspecification and Noise Level Assumption}\label{sec:analytical_II}
We now present our second main result, Theorem~\ref{thm:emse_noise_model} which generalizes the setting of Theorem~\ref{thm:emse_noisfree} by allowing the assumed noise level  to be non-zero.
\begin{theorem}\label{thm:emse_noise_model}
    Let $\Khatxbar = \eye{\pbar}$, $\Kv=\sigmav^2\eye{n}$, 
    $\Khat_{\vbar} = \sigmahatvbar^2\eye{n} \succ 0$,
    and $\pbar>1$.
    Then the expected MSE associated with $\Wbar$ in \eqref{eqn:lmmse_false} 
    is
    \begin{align}\begin{split}\label{eqn:thm2:emse}
        \egen(p_S,p_C,p_F,n)
        & = (\tr(\KxC) + \sigmav^2)\frac{p_S}{\pbar}\muM_1 \\
        &\qquad +\muM_2 \tr(\KxS) + \tr(\KxC),
    \end{split}
    \end{align}
    where 
    \begin{equation}\label{eqn:thm:muM_1}
        \muM_1 = \sum_{i=1}^{\pbar} \eunder{\lambda_i}\Big[
                \frac{\lambda_i}{(\lambda_i + \sigmahatvbar^2)^2}
            \Big],
    \end{equation}
    and
    \begin{align}\begin{split}\label{eqn:thm:muM_2}
        \!\! \muM_2 
        & = \frac{1}{\pbar(\pbar+2)}\Big((p_S+2)\sum_{i=1}^{\pbar} 
        \eunder{\lambda_i}\Big[
            \frac{\sigmahatvbar^4}{(\lambda_i + \sigmahatvbar^2)^2}
        \Big] \\
            &
        + 2\frac{\pbar - p_S}{\pbar-1}
        \sum_{i=1}^{\pbar}   \sum_{j=1}^{i-1} 
        \eunder{\lambda_i, \lambda_j}\Big[
            \frac{\sigmahatvbar^4}
            {(\lambda_i + \sigmahatvbar^2)(\lambda_j + \sigmahatvbar^2)}\Big]
        \Big),
    \end{split}
    \end{align}
    and $\lambda_i$ are the eigenvalues of $\Abar\T\Abar\in\Rbb^{\pbar\times\pbar}$.
        
\end{theorem}

\noindent Proof: See Appendix~\ref{proof:thm:emse_noise_model}.
Note that while the setting of Theorem~\ref{thm:emse_noisfree} is a special case of Theorem~\ref{thm:emse_noise_model}
in the limit of $\sigmahatvbar\rightarrow0$,
we have kept the results separate 
since the setting with $\sigmahatvbar=0$ allows  more explicit evaluations of the expressions. 

\subsection{Effect of Fake Features and Noise Level Assumption}
    \label{sec:analytical_II:interplay_regularizers}
    The expressions in Theorems~\ref{thm:emse_noisfree} and \ref{thm:emse_noise_model} show how the presence of fake features can have a regularizing effect on the expected MSE.
    Additionally in Theorem~\ref{thm:emse_noise_model},
    we see the regularizing effect of the noise level assumption $\sigmahatvbar$.
    We will now discuss these effects in detail.

    In the setting of Theorem~\ref{thm:emse_noise_model},
     the potentially misspecified covariance matrices are given by 
    $\Khatxbar = \eye{\pbar}$ and
    $\Khat_{\vbar} = \sigmahatvbar^2\eye{n} $.
    By \eqref{eqn:lmmse_false},
    with these covariance matrices,
    the estimator is
    $\Wbar = \Abar\T(\Abar\Abar\T + \sigmahatvbar^2\eye{n})\inv$,
    which can be rewritten as
    \begin{equation}\label{eq:Wbar:ATAform}
        \Wbar=(\Abar\T\Abar + \sigmahatvbar^2\eye{\pbar})^{-1} \Abar\T.
    \end{equation}
    Hence, 
    if $\Abar\T\Abar + \sigmahatvbar^2\eye{\pbar}$ is ill-conditioned 
    then it will affect the behaviour of the estimator. 
    
    Recall that $\Abar \in \Rbb^{n \times \pbar}$,
    where $n$ is the number of observations,
    and $\pbar = p_S+p_F$ is the number of unknowns in the misspecified model.
    It has been established that the singular values of an $n\times\pbar$ matrix with i.i.d. zero-mean random entries with unit variance lie on the interval 
    $[\sqrt{n} - \sqrt{\pbar}, \sqrt{n} + \sqrt{\pbar}]$
    with high probability \cite{rudelson_non-asymptotic_2010}. 
    Asymptotically as $n$ and $\pbar$ grows,
    all singular values lie in this interval.
    The non-zero eigenvalues $\lambda_i$ of $\Abar\T\Abar$ are the squared singular values of $\Abar$,
    hence these $\lambda_i$ are lower bounded by
    $\ell_{\min} \triangleq (\sqrt{n} - \sqrt{\pbar})^2$,
    with high-probability.
    Now suppose that the assumed noise level $\sigmahatvbar$ is small in relation to $\ell_{\min}$,
    and note that $\lambda_i$ appears in the denominators of the fractions $\frac{\lambda_i}{(\lambda_i+\sigmahatvbar^2)^2}$ in \eqref{eqn:thm:muM_1}.
    These fractions are then $\approx\frac{1}{\lambda_i}$,
    which can take on very large values if $\ell_{\min}$ is close to zero,
    i.e., 
    if $n\approx \pbar$.
    
    \begin{remark}
        With $\sigmahatvbar^2$ small,  the peak in MSE when the number of samples is close to the assumed model size,
        i.e., $n\approx \pbar$,
        occurs because if $n\approx \pbar$, then non-zero eigenvalues of $\Abar\T\Abar$ may be close to zero (but not exactly zero) with high probability. 
        By changing the number of fake features $p_F$,
        the probability of having non-zero eigenvalues close to zero decreases,
        hence the problem becomes effectively regularized.
    \end{remark}

    Similarly,
    the MSE in Theorem~\ref{thm:emse_noisfree} (where $\sigmahatvbar^2=0$) also takes on large values if $n\approx\pbar$,
    where the effect of the dimensions on the error can be directly seen in $n-\pbar-1$ or $\pbar-n-1$,
    which appear in the denominators of the terms in \eqref{eqn:thm:emse_noisefree:n_geq_pbar} and \eqref{eqn:thm:emse_noisefree:n_less_pbar}.
    Hence, $p_F$ can act as a regularizer  both with $\sigmahatvbar>0$ under small $\sigmahatvbar$ and with $\sigmahatvbar=0$.

    Nevertheless, 
    the MSE peak at $n \approx \pbar$ can be dampened by a large enough $\sigmahatvbar$.
    In particular, consider the fractions $\frac{\lambda_i}{(\lambda_i+\sigmahatvbar^2)^2}$ in \eqref{eqn:thm:muM_1}.
    If $\sigmahatvbar$ is large enough in relation to the eigenvalue distribution's lower bound $\ell_{\min}$,
    then these fractions take small values with high probability, preventing divergent error behaviour.
    Although  $\sigmahatvbar$ can be used to regularize the problem and  dampen the peak in MSE around $n\approx \pbar$,
    its value should be not be  too high.
    The next remark illustrates this point: 
    \begin{remark}
        In the setting of Theorem 2,
        the expected MSE is constant in the limit of $\sigmahatvbar\rightarrow\infty$,
        for any $n<\infty$, $\pbar<\infty$:
        \begin{equation}\label{eqn:limit:sigmahat_infty}
            \lim_{\sigmahatvbar\rightarrow\infty} \egen(p_S,p_C,p_F,n)
            = \tr(\Kx).
        \end{equation}

    \end{remark}
    This result is a straightforward consequence of the fact that  $\Wbar\rightarrow\bm 0$ in \eqref{eqn:lmmse_false} as $\sigmahatvbar\rightarrow\infty$. 
    Note that $\tr(\Kx)$ is the a priori uncertainty for $\xvec$, hence \eqref{eqn:limit:sigmahat_infty} shows that when $\sigmahatvbar$ is too high,
    little or no reduction in uncertainty is gained with estimation.

    We now discuss $\Wbar$ in relation to the regularized least-squares (LS) approach
    \begin{equation}
        \hat{\xbar}_{LS} 
        = \arg \min_{\xbar} \| \yvec - \Abar \xbar\|^2 +\mu \| \xbar \|^2,
    \end{equation}
    where $\mu > 0$ is the regularization parameter.
    This framework is typically referred to as ridge regression,
    or Tikhonov regularization \cite{b_TikhonovArsenin},
    and has been well-studied.
    Recent works have focused on the perspective of double descent \cite{nakkiran2020optimal},
    and robust estimation \cite{MartinezCamaraMumaZoubirVetterli_2015,SulimanBallalKammounAlNaffouri_2016}.
    The optimal LS solution is given by
    $\hat{\xbar}_{LS} = \Wbar \yvec$,
    with $\Wbar$ from \eqref{eq:Wbar:ATAform},
    and $\sigmahatvbar^2 =\mu>0$.
    The regularization term with $\mu > 0 $ is known to mitigate effects of the potentially ill-conditioned matrix $\Abar$,
    confirming the role of $\sigmahatvbar^2$ as a regularization parameter.
    As discussed above, 
    our results illustrate that $p_F$ plays a similar regularizing role. 

\kern-0.25em
\subsection{Optimal Noise Level Assumption}
\kern-0.25em
We will now consider the special case with a large number of observations, i.e., $n\gg \pbar$, and present the optimal $\sigmahatvbar$ that minimizes the expected MSE.

\begin{corollary}\label{col:n_gg_pbar}
Consider the setting of Theorem~\ref{thm:emse_noise_model}.
If $n\gg\pbar$,
    then the expected MSE is 
    \begin{align}\label{eqn:approx_n_gg_pbar}
    \begin{split}
        \egen(p_S,p_C,p_F,n) 
        &\approx 
        (\tr(\KxC) + \sigmav^2)\frac{p_S}{\pbar}\muMtilde_1 \\
        & \qquad + \muMtilde_2 \tr(\KxS) + \tr(\KxC),
    \end{split}
    \end{align}
    with
    \begin{equation}\label{eqn:muM_1_and_2_n_gg_pbar}
        \muMtilde_1 = \frac{n\pbar}{(n + \sigmahatvbar^2)^2},
        \quad \muMtilde_2 = \frac{\sigmahatvbar^4}{(n + \sigmahatvbar^2)^2}.
    \end{equation} 

\end{corollary}

\noindent Proof: See Appendix~\ref{proof:col:n_gg_pbar}.

Corollary~\ref{col:n_gg_pbar} gives an approximation of the expected MSE for settings where there is a high number of samples $n$ in relation to the number of unknowns in the misspecified model $\pbar$.
We further note that,
for $\sigmahatvbar<\infty$ and $\pbar < \infty$,
we have 
\begin{equation}\label{eqn:thm2:egen_limit_n_infty}
    \lim_{n\rightarrow\infty} \egen(p_S,p_C,p_F,n) = \tr(\KxC).
\end{equation}
    
By analyzing \eqref{eqn:muM_1_and_2_n_gg_pbar},
we see how the expected MSE $\egen$ is affected by the assumed noise level $\sigmahatvbar$.
The following result gives the value of $\sigmahatvbar^2$ which minimizes $\egen$.
\begin{lemma}\label{lem:optimal_sigmahat}
    Consider the setting of Theorem~\ref{thm:emse_noise_model},
    and the expression
    \begin{equation}\label{eq:optimal_sigmahat}
        \sigmahatvbaropt^2 = p_S \frac{\tr(\KxC) + \sigmav^2}{\tr(\KxS)}.
    \end{equation}
    If $p_F=0$,
    then the $\sigmahatvbar^2$ that minimizes the expected MSE is
    \begin{equation}
        \arg \min_{\sigmahatvbar^2} \egen(p_S,p_C,p_F,n) 
        = \sigmahatvbaropt^2.
    \end{equation}
    If $p_F>0$ and $n\gg\pbar$,
    then 
    \begin{equation}\label{eqn:optimalsigmabar:nggpbar}
        \arg \min_{\sigmahatvbar^2} \egen(p_S,p_C,p_F,n) 
        \approx \sigmahatvbaropt^2.
    \end{equation}
    
\end{lemma}

\noindent Proof: See Appendix~\ref{proof:lem:optimal_sigmahat}.

Note that in general, $\sigmahatvbaropt$ in \eqref{eq:optimal_sigmahat} is not equal to $\sigmav$. Instead, 
$\sigmahatvbaropt$ can be interpreted as the effective noise level of the misspecified model.
For instance,
if $\tr(\KxC)$ or $\sigmav^2$ is large in comparison to $\tr(\KxS)$,
then the features that are included in the model through $\AS$ can not explain, 
i.e., account for, 
a large portion of $\yvec$.
Hence, $\sigmahatvbaropt^2$ increases with $\tr(\KxC)$.

In general, we expect the number of fake features $p_F$ to affect the optimal $\sigmahatvbar$.
However, there is no such effect in \eqref{eqn:optimalsigmabar:nggpbar} where $n \gg \pbar$.
While finding the optimal $\sigmahatvbar$ for a general $p_F$ remains an important line of future work,
we illustrate how the optimal $\sigmahatvbaropt$ changes with $p_F$ in our numerical results in Section~\ref{sec:numerical:nonzero_sigmahatvbar}.

\balert{%
\section{Expected MSE for Predicting the Observations}\label{sec:LabelMSE}

Up to now, we have focused on the MSE associated with the unknowns $\xS$ and $\xC$. 
We now consider the output error, 
i.e., the error when predicting the output $y_*$ associated with the  pair $(y_*, \avec_*)$. 
In particular, 
    the \textit{output MSE},
    i.e., the error related to the estimator $\Wbar$ and a data pair 
    $(y_*, [\avec_{S*}\T,\avec_{C*}\T,\avec_{F*}\T]\T)$ unseen during training,
    is given as
    \begin{equation*}
        J_y(\Wbar) 
        = \Ebb_{y_*, \xvec, \yvec}[ (y_* - \abar_*\T \xbarhat)^2]
        = J(\Wbar) + J_F(\Wbar_F) + \sigmav^2,
    \end{equation*}
    with $J(\Wbar)=J_S(\WbarS)+ J_C(\WbarC)$ as in \eqref{eqn:mse_WbarC_zero},
    $\Wbar_F$ as in \eqref{eqn:lmmse_false},  $J_F(\Wbar_F)$ as in \eqref{eqn:JF_WF}
    and 
    $\abar_* = [\avec_{S*}\T,\avec_{F*}\T]\T$,
    $\xbarhat = [\xShat\T, \xFhat\T]\T$, 
    and $y_* = \atilde_*\T \xvec + v_*$,
    where
    $\atilde_* = [\avec_{S*}\T,\avec_{C*}\T]\T$ 
    and $\xvec = [\xS\T,\xC\T]\T$.

    Taking the expectation of the output MSE $J_y(\Wbar)$ over the distribution of $\Amat$,
    the \textit{expected output MSE} is defined as
    \begin{align}
        \egen_y = \eunder{\Amat}[J_y(\Wbar)] 
        & = \egen_S + \egen_C + \egen_F + \sigmav^2 \label{eqn:egen_y=egen_S+egen_C+egen_F+sigma^2},
    \end{align}
    where $\egen_S = \Ebb_{\Amat}[J_S(\WbarS)]$,
    $\egen_C = \Ebb_{\Amat}[J_C(\WbarC)]$,
    and  $\egen_F$ is the error associated with the fake features $\AF$ and the estimate $\xFhat$,
    i.e.,
    \begin{equation}\label{eqn:egen_F_def}
        \egen_F(p_S, p_C, p_F, n) = \eunder{\Amat}[ J_F(\Wbar_F) ].
    \end{equation}
 
    \begin{theorem}\label{thm:egen_F}
        Consider the setting in Theorem~\ref{thm:emse_noisfree}.
        If $n > \pbar + 1$,
        then 
        \begin{equation}
            \egen_F(p_S, p_C, p_F, n)
            = \tfrac{p_F}{n - \pbar - 1}(\tr(\KxC) + \sigmav^2).
        \end{equation}
        If $\pbar > n + 1 $,
        then
        \begin{align}
        \begin{split}
            \egen_F(p_S, p_C, p_F, n)
            = & \tfrac{n p_F}{\pbar(\pbar - n - 1)} 
            ( \tr(\KxC) + \sigmav^2) \\
            & \quad + \tfrac{n p_F(\pbar - n)}{(\pbar - 1)\pbar(\pbar + 2)}\tr(\KxS).
        \end{split}
        \end{align}
        
    \end{theorem}
    
    \noindent Proof: See Appendix~\ref{sec:proof:thm:egen_F}.

    Inserting the expressions for $\egen=\egen_S+\egen_C$ in Theorem~\ref{thm:emse_noisfree} 
    and $\egen_F$ in Theorem~\ref{thm:egen_F} 
    into \eqref{eqn:egen_y=egen_S+egen_C+egen_F+sigma^2}
    we obtain that if $n > \pbar + 1$,
    then the expected output MSE is
    \begin{equation}\label{eqn:egen_y_decomposed}
        \egen_y
        = \frac{\pbar }{n - \pbar - 1}( \tr(\KxC) + \sigmav^2) + \tr(\KxC) + \sigmav^2,
    \end{equation}
    and if  $\pbar > n + 1$,
    then
    \begin{align}\begin{split}
        \egen_y
        &= \frac{n}{(\pbar - n - 1)}(\tr(\KxC) + \sigmav^2) \\
        & \qquad + \big(1 - \frac{n}{\pbar}\big)\tr(\KxS) + \tr(\KxC) + \sigmav^2.
    \end{split}
    \end{align}
    }
    \balert{%
    The power of the missing unknowns appears in the output MSE together with the noise level,
    i.e.,
    ${\tr(\KxC) + \sigmav^2}$.
    This is consistent with the fact that from the perspective of the misspecified model,
    the missing signal and the inherent noise can be together regarded as an effective noise term.
    }

    \balert{%
    Here, $\egen_y$ is consistent with \cite[Theorem 2.1]{belkin_two_2019},
    with the change of variables 
   $\|\beta_{T^c}\|^2 \to \tr(\KxC)$ and  $\|\beta_T\|^2 \to \tr(\KxS)$. Note that Thm.~\ref{thm:emse_noisfree} and Thm.~\ref{thm:egen_F} provide a non-trivial decomposition of this error that has not been studied in the literature.
    In Section~\ref{sec:numerical-decomposition},
    we investigate the decomposition in \eqref{eqn:egen_F_def} numerically.
    Interestingly,
    our results there illustrate that $\xShat$ can have relatively low error,
    even though the output MSE is above its asymptote of $\tr(\Kx)+\sigmav^2$ as $p_F\rightarrow\infty$.

    }

    \kern-0.125em
\section{Numerical Results}\label{sec:numerical}
\kern-0.125em

\balert{%
\subsection{Example with Liver Toxicity Data}
\label{sec:practical_example}
}
\balert{%
We now illustrate the double-descent behaviour with real-world data using the \textit{liver toxicity} dataset available in the \texttt{mixOmics} package \cite{Rohart_2017_mixOmics},
containing measurements of toxin levels in blood samples from $64$ rats.
We estimate the level of a toxin (urea nitrogen) related to liver injury,
using genetic data of $3116$ genes.
We perform $M=1000$ experiments,
for which we  choose $n=54$ of the data points uniformly at random for training,
and use the remaining 
$n_*=10$ to compute the empirical output MSE,
i.e., the error in $y$, see \eqref{eqn:egen_y_decomposed}.
For each experiment, 
we increase the number of features $\pbar$ used for estimation,
such that $\pbar=1,\,\dots,\,3116$,
and record the output MSE.
Hence for experiment $(i)$,
$i=1,\,\dots,\,M$,
we have the training data as 
$\Abar^{(i)}\in\Rbb^{n\times \pbar}$ and 
$\yvec^{(i)}\in\Rbb^{n\times 1}$,
and estimate the unknowns as 
$\xbarhat^{(i)} = \Abar^{{(i)T}}(\Abar^{(i)}\Abar^{(i)T} + \sigmahatvbar^2\eye{\pbar})\p \yvec^{(i)} $,
and then compute the output MSE on the $n_*$ unseen data as $\frac{1}{n_*}\|\yvec_*^{(i)} - \Abar_*^{(i)} \xbarhat^{(i)}\|^2$,
where $\yvec_*^{(i)}\in\Rbb^{n_*\times 1}$ and $\Abar_*^{(i)}\in\Rbb^{n_*\times \pbar}$.

We plot the empirical average of the output MSE over $M$ experiments versus the number of observed features $\pbar$ for four different noise level assumptions $\sigmahatvbar$ in Figure~\ref{fig:liver}.
We observe that for small $\sigmahatvbar$,
the output MSE exhibits a double-descent behaviour over $\pbar$,
with its peak in error around the threshold $\pbar=n$.
The four curves show that the output MSE is minimized  with $\pbar$ that is much larger than the number of training samples,
i.e., $n=54$.
%
Hence,
the lowest output MSE over all $\pbar$ can be obtained in the overparametrized regime.

\begin{figure}
    \centering
    \includegraphics{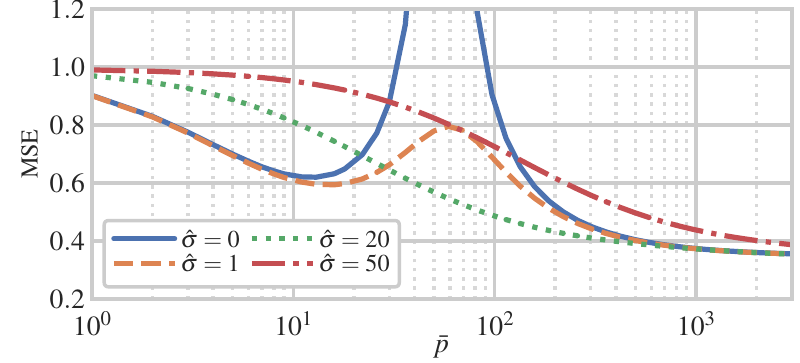}
    \caption{
    \balert{%
    The average prediction error versus the number of features $\pbar$ for \textit{liver toxicity} dataset  \cite{Rohart_2017_mixOmics}.
    }
    }
    \label{fig:liver}
\end{figure}

}

\subsection{Experimental Setup and Overview}
\kern-0.25em
We now provide the setting for the rest of the numerical results. 
The numerical results are obtained by averaging over $M_r=100$ realizations $(i)$ of the regressors and $M_u=100$ realizations $(j)$ of the unknowns and noise for each realization of the regressors.
We draw one set of regressors $\Amat^{(i)}=[\AS^{(i)}, \AC^{(i)}, \AF^{(i)}]$
from the standard Gaussian distribution for $(i)$, $i=1,\dots,M_r$.
For each set of regressors $\Amat^{(i)}$,
we draw one vector of unknowns $\xvec^{(i,j)}$ from 
$\Nc(0,\Kx)$
with $\Kx=\eye{p}$,
and one noise vector $\vvec^{(i,j)}$ from $\Nc(0,\sigmav^2\eye{n})$.
The observations $\yvec^{(i,j)}$ are generated using the underlying system \eqref{eqn:model_true},
and the estimator $\Wbar^{(i)}$ is computed as in \eqref{eqn:lmmse_false},
based on the misspecified model in \eqref{eqn:model_false},
with $\Khatxbar = \eye{\pbar}$, 
$\Khatvbar = \sigmahatvbar^2 \eye{n}$.
Then, the estimate $\xbarhat^{(i,j)} = \Wbar^{(i)} \yvec^{(i,j)} $ is computed.
The MSE $\Jtilde^{(i)}(\Wbar^{(i)})$ is then computed as 
\setlength{\belowdisplayskip}{1pt}    
\begin{equation*}
    \Jtilde^{(i)}(\Wbar^{(i)}) 
    \!=\! \frac{1}{M_u}\!\sum_{j=1}^{M_u} \left(
            \|\xS^{(i,j)} \! - \! \xShat^{(i,j)}\|^2
            \! + \! \|\xC^{(i,j)} - \xChat^{(i,j)}\|^2
        \right)\!,
\end{equation*}
with $\xChat^{(i,j)}= 0$.
$\Jtilde^{(i)}(\Wbar^{(i)})$ is then averaged over the $M_r$ realizations of regressors,
to create the empirical average MSE,
as an estimate of the expected MSE $\egen$ in \eqref{eqn:emse_false} 
\begin{equation}\label{eqn:mse_empirical}
    \hat{\egen}(p_S, p_C, p_F, n) 
    \triangleq \frac{1}{M_r} \sum_{i=1}^{M_r} \Jtilde^{(i)}(\Wbar^{(i)}).
\end{equation}
\setlength{\belowdisplayskip}{4pt}    
In the plots, we report the \textit{normalized MSE} given by $\frac{\hat{\egen}}{\tr(\Kx)}$. 

    In the case of $\sigmahatvbar\!= \! 0$,
    analytical curves obtained using  $\egen$ in Theorem~\ref{thm:emse_noisfree}.
    For $\sigmahatvbar > \! 0$,  
    analytical curves are obtained 
    using Theorem~\ref{thm:emse_noise_model}
    and numerical integration to obtain the necessary moments  \cite[Section~1.2]{TulinoVerdu_2004}.

We now conduct a series of experiments focusing on how the expected MSE $\egen$,
and its empirical counterpart $\hat \egen$,
depends on $n$, $p_F$, $\sigmahatvbar$ and $\sigmav$ 
In all plots, the lines represent the empirical results and the markers represent the analytical results. 
We observe that in all applicable cases there is a close match between the empirical and analytical curves.

\kern-0.25em
\subsection{Effect of Fake and Missing Features}
\kern-0.25em
    \begin{figure}[t]
        \centering
        \includegraphics[width= \linewidth]{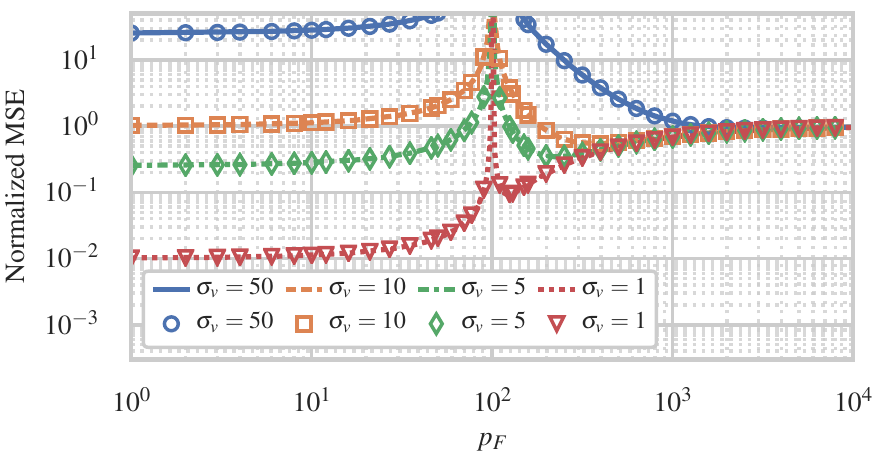}
        \caption{
            Empirical (lines) and analytical (markers) MSE versus the number of fake features $p_F$.
            Here, $p_S = 100$, $p_C=0$,
            $\sigmahatvbar=0$ and $n=200$.
        }
        \label{fig:fig3}
    \end{figure}

    \begin{figure}[t]
        \centering
        \includegraphics[width=\linewidth]{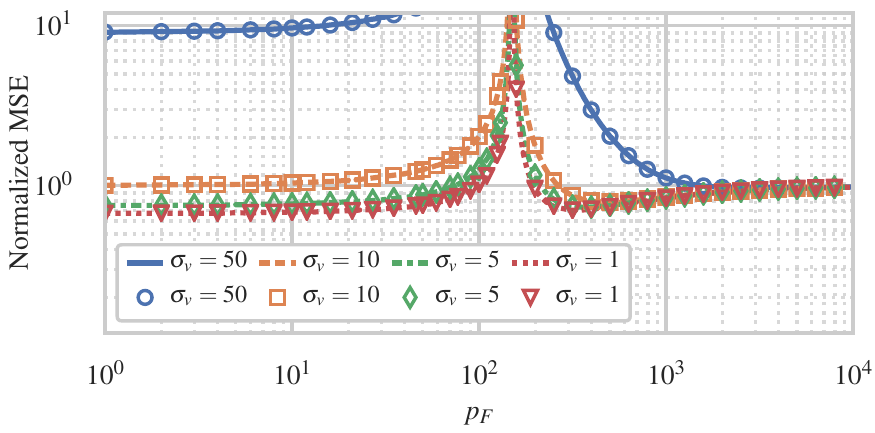}
        \caption{
            Empirical (lines) and analytical (markers) MSE versus the number of fake features $p_F$.
            Here, $p_S = 50$, $p_C=50$,
            $\sigmahatvbar=0$ and $n=200$.
        }
        \label{fig:fig4}
    \end{figure}
    
    We now illustrate the effect of the number of fake features $p_F$,
    and how the number of missing features $p_C$ affects this relationship.
    
    In Figures~\ref{fig:fig3} and \ref{fig:fig4}, 
    we plot the theoretical expected MSE from Theorem~\ref{thm:emse_noisfree},
    and the empirically averaged MSE against $p_F$ in the misspecifed model.
    In Figure~\ref{fig:fig3},
    there are no missing features, i.e., $p_C=0$,
    and in Figure~\ref{fig:fig4},
    we have $p_C=50$.
    The MSE is plotted for four different noise levels: $\sigmav=\{1,5,10,50\}$,
    and the noise level assumption of $\sigmahatvbar=0$.
    Here, $n=200$,
    and recall that $\pbar=p_S + p_F$, 
    hence the threshold $n = \pbar$ corresponds to $p_F=100$ in Figure~\ref{fig:fig3} ($p_S=100$) 
    and $p_F=150$ in Figure~\ref{fig:fig4} ($p_S=50$).

    As discussed in Section~\ref{sec:analytical}, 
    we observe that the optimal $p_F$ is not always $p_F=0$,
    i.e., the presence of the fake features can improve the estimation performance.
    As $p_F$ increases from $p_F=1$,
    the MSE increases more and more rapidly as $\pbar = p_S + p_F$ approaches $n$.
    If $p_F$ increases further, then the MSE decreases until it hits a local minimum and then increases and eventually converges to $\egen = \tr(\Kx)$.
    Although all curves converge to the same value, i.e., $\tr(\Kx)$,
    the local minima in the region $p_F>n$ can be well below $\tr(\Kx)$,
    and comparable to the MSE for small $p_F$.
    For instance,
    for the curves in Figure~\ref{fig:fig3} and \ref{fig:fig4} with $\sigmav=10$,
    the local minima around $p_F=400$ are lower than the minimum MSE for smaller $p_F$.
    If the noise is even larger at $\sigmav=50$,
    then the MSE is very high for small $p_F$,
    and significantly lower for $p_F>1000$,
    and still approaches $\tr(\Kx)$ as $p_F\rightarrow\infty$.
    Hence, these results illustrate the fake features' regularizing effect.

    Effects of missing features can be seen by comparing Figure~\ref{fig:fig3} and \ref{fig:fig4}. 
    In Figure~\ref{fig:fig3},
    where $p_C=0$, 
    we observe that
    the MSE for small $p_F$ scales with the noise level $\sigmav^2$
    (recall that the y-axis is normalized by $\tr(\Kx)$).
    On the other hand, 
    in Figure~\ref{fig:fig4},
    the MSE for small $p_F$ scales with the ``effective'' noise level, i.e. the power of the unobserved unknowns $\tr(\KxC)$,
    together with the noise level $\sigmav^2$. 

\kern-0.25em
\subsection{Effect of Non-zero Noise Level Assumption}
\kern-0.25em
    \label{sec:numerical:nonzero_sigmahatvbar}

\begin{figure}[t]
        \centering
        \includegraphics[width=\linewidth]{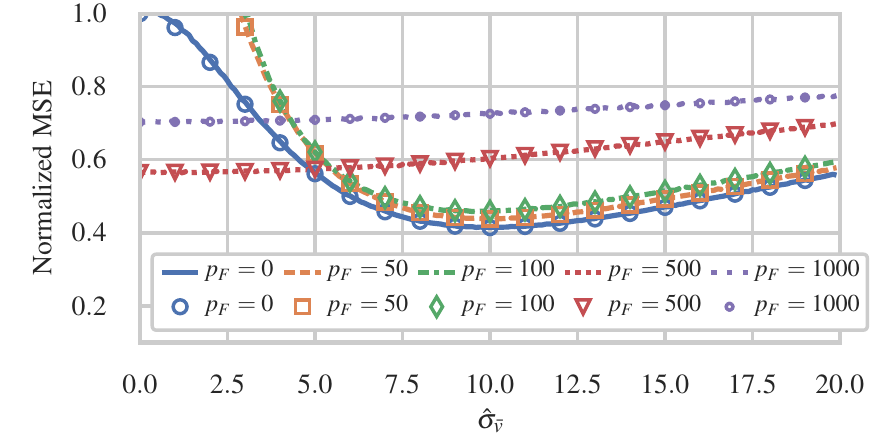}
        \caption{
            The expected and average MSE versus the assumed noise level $\sigmahatvbar$.
            Here $p_S=100$, $p_C=0$, $\sigmav=10$ and $n=200$.
        }
        \label{fig:fig12}
    \end{figure}

    We now investigate the effect of having a non-zero noise level assumption,
    i.e., $\sigmahatvbar\neq 0$.
    Recall that the assumed noise level $\sigmahatvbar$ is not the same value as the noise level $\sigmav$ of the underlying system.
    In Figure~\ref{fig:fig12}, we plot the MSE versus $\sigmahatvbar$ for different values of $p_F$,
    with $\sigmav=10$.
    Here, $n=200$, $p_S=100$ and $p_C=0$.
        
    In previous figures,
    where $\sigmahatvbar=0$,
    we observe large peaks in MSE if $n=\pbar$.
    If $\sigmahatvbar$ is large enough,
    such a peak can be damped as can be observed in Figure~\ref{fig:fig12},
    where the threshold occurs if $p_F=100$.
    In particular,
    we observe that if $\sigmahatvbar>5$ the respective MSE for $p_F=0$ and $p_F=100$ are close in value,
    compared to when $\sigmahatvbar<5.0$.
    
    Figure~\ref{fig:fig12} illustrates
    that the optimal choice of $\sigmahatvbar$ varies with $p_F$.
    For example, 
    if $p_F=0$,
    then the optimal $\sigmahatvbar=10$,
    and if instead $p_F=500$,
    then the optimal $\sigmahatvbar\approx 0$.
    Hence we observe a trade-off between the regularizing effects provided by the fake features and by $\sigmahatvbar$.
    However,
    note that the optimal value $\sigmahatvbaropt=10$ is given by Lemma~\ref{lem:optimal_sigmahat} for $n>>p_F$
    performs quite well over $0<p_F\leq 100$ here.
    
\balert{%
    \subsection{The Output MSE and its Decomposition}
    \label{sec:numerical-decomposition}
}
\begin{figure}
    \centering
    \includegraphics{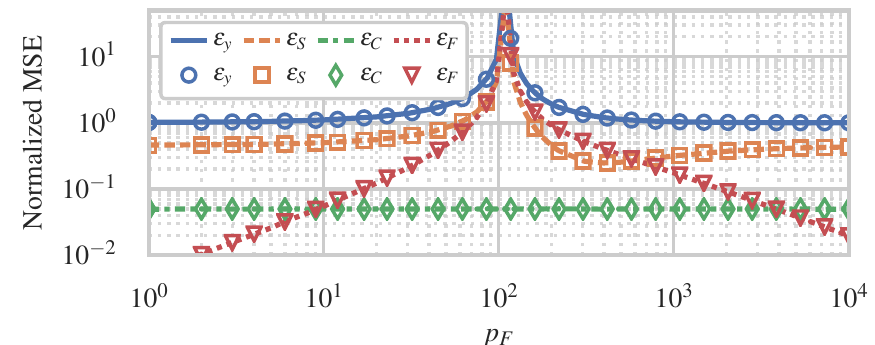}
    \caption{\balert{%
    Empirical (lines) and analytical (markers) output MSE $\varepsilon_y$ and its decomposition,
    versus the number of fake features $p_F$.
    Here, $p_S = 90$, $p_C=10$, $\sigmav=10$, $\sigmahatvbaropt=0$ and $n=200$.}}
    \label{fig:fig17}
\end{figure}

    \balert{%
        In Figure~\ref{fig:fig17}, we plot the expected output MSE $\egen_y$,
        i.e., the error in $y$, see \eqref{eqn:egen_y_decomposed},
        together with its components $\egen_S$, $\egen_C$, $\egen_F$, see \eqref{eqn:egen_y_decomposed}.
        This figure highlights that the output MSE and the MSE in the unknowns can behave drastically different:}
    \balert{%
    In particular, 
    even when the MSE associated with $\yvec$ stays on the same level for the over/under parametrized scenarios (except around the peak),  the minimum error for the unknowns $\xS$ can be obtained in the overparametrized case with a high number of fake features.
    The plots also illustrate that the estimate for the unknowns $\xS$ can be of relatively high quality,
    even though the output MSE is high. 
    In other words,
    these results suggest that the model may learn the parameters $\xS$ of the underlying system, 
    which is a subset of the true parameters, 
    even though the prediction for the output is poor.
}

\kern0.5em
\balert{%
\subsection{The Effect of the Data Covariance}\label{sec:covariance-numerical}
}
\kern-0.5em

\begin{figure}
    \centering
    \includegraphics{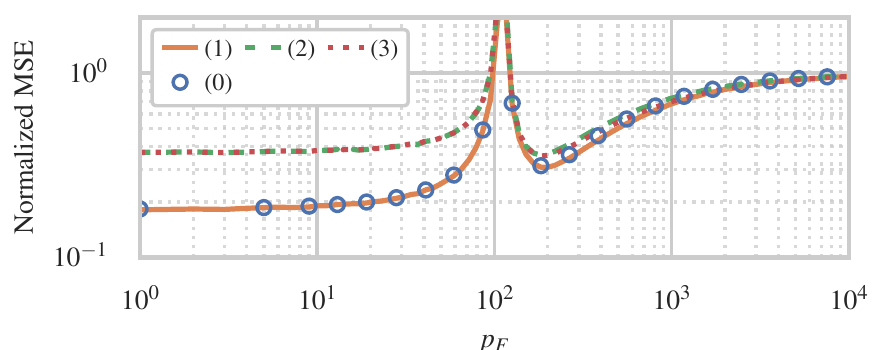}
    \caption{\balert{The MSE versus the number of fake features for experiments \textit{(0) -- (3)} which consider different feature covariances.
    Here, $p_S=90$, $p_C=10$, $\sigmav=0.1$, $\sigmahatvbar=0$ and
    $n=200$.}}
    \label{fig:fig20}
\end{figure}

\balert{%
We now investigate the effect of non-isotropic features. 
Let $\Ka$ and $\Kaf$ be the feature covariance matrices  for $\Amat$ and $\AF$ so that  each row is i.i.d. drawn from $\Nc(0, \Ka)$ and $\Nc(0, \Kaf)$,
respectively.
Here, $\Amat$ and $\AF$ are uncorrelated. 
We generate $\Ka$ as follows:
\textit{i)} Given a decay parameter $\alpha$, generate a matrix of eigenvalues $\Lambdamat_a = \diag(1^\alpha, \dots, p^\alpha)\in\Rbb^{p\times p}$;
\textit{ii)} Generate a Haar distributed orthogonal matrix $\Umat_{a} \in\Rbb^{p\times p}$ 
\cite{anderson_generation_1987};
\textit{iii)} Set $\Ka = \frac{p}{\tr(\Lambdamat_a)}\Umat_a\Lambdamat_a\Umat_a\T$.
The matrix $\Kaf$ is generated with the same procedure using a different decay parameter $\alpha_F$ for 
$\Lambdamat_{a_F}\in\Rbb^{p_F\times p_F}$ and an independently generated orthogonal matrix 
$\Umat_{a_F}\in\Rbb^{p_F\times p_F}$.
We perform the  experiments with the following pairs of $(\alpha,\alpha_F)$:
\textit{(0)} $\alpha=0$, $\alpha_F=0$;
\textit{(1)} $\alpha=0$, $\alpha_F=20$;
\textit{(2)} $\alpha=1$, $\alpha_F=0$;
\textit{(3)} $\alpha=1$, $\alpha_F=20$.
Here, $\Kx = \eye{p}$ $p_S = 90$, $p_C=10$, $n=200$ and $\sigmav=10$.

In Figure~\ref{fig:fig20},
we plot the simulated average MSE (lines) in the unknowns
versus the number of fake features $p_F$ for the experiments \textit{(1)-(3)}.
Experiment \textit{(0)} fulfills the setting of Theorem~\ref{thm:emse_noisfree},
hence we plot the analytical expected MSE.
We observe two main effects of having non-isotropic features in the underparameterized regime where $n\geq \pbar$:
Firstly, for a given $\Ka$,
changing $\Kaf$ does not change the average MSE.
This can be seen by comparing the curves for experiment \textit{(0)} and \textit{(1)},
and for \textit{(2)} and \textit{(3)}.
Secondly, we observe that the MSE increases with $\alpha$,
or in other words the MSE is higher for more correlated underlying features.
In the overparameterized regime,
i.e., $n<\pbar$,
the curves are close,
suggesting less dependence on the correlation structure.}

    \kern0.5em
\section{Discussions}\label{sec:discussions}
\kern-0.5em
    We have shown that fake features can decrease the error and hence improve the estimation performance, 
    even though these features are uncorrelated with the true features. 
    \balert{%
    Under the Gaussian feature assumptions in Section~\ref{sec:problem:model_and_misspec},   we have shown that if a performance improvement is observed with a larger dictionary, 
    this improvement is not necessarily due to the explanatory power of the added features;
    it can be merely due to the regularization effect of fake features.}
    We note that this result is a consequence of the misalignment between the true model and the assumed model,
    which is a typical situation in practical scenarios. 

\kern-0.15em
\section{Conclusions}\label{sec:conclusions}
\kern-0.15em
\balert{
    We have proposed a model misspecification framework 
    which enables a joint treatment of fake features,
    together with incorrect covariance assumptions on the unknowns and the noise.}
    We have revealed the trade-offs between the fake features and the noise level assumptions,
    when data comes from an underlying linear system
    and the estimator is based on a misspecified linear  model.
    
    Based on LMMSE estimation, our main results  provide analytical expressions of the expected MSE of the misspecified estimator.    
    We have presented numerical experiments which verify our analytical findings.
    Our results show that the presence of fake features can improve the estimation performance even when the model is overparameterized.
    We illustrate that the MSE exhibits double descent with increasing model size, 
     even when the model size increase is due to fake features.
    \balert{Our results further show that even though the respective MSEs associated with the model parameters  and the prediction of the output both exhibit double descent, they can have significantly different behaviour with increasing model size.}
    \balert{In particular,
    the globally optimal number of fake features for the MSE in the model parameters may be found in the overparameterized regime,
    even though the output MSE remains high regardless of the number of fake features.}
 
 \balert{
    Important directions for future work include extending the error characterization to general regressor and misspecified model structures, including arbitrary covariance models,  as well as model mismatch between non-linear and linear models. }

    \setlength{\abovedisplayskip}{2pt}
\setlength{\belowdisplayskip}{2pt}
\kern-1em
\appendix
\section{Appendix}
\kern-0.5em
\subsection{Preliminaries}
\kern-0.25em
    We here provide an overview of results which are used throughout the derivations of the main results of our paper.
    
    \begin{lemma}\label{lemma:egen}
        Consider the setting in Section~\ref{sec:problem}.
        Let $\Kv = \sigmav^2\eye{n}$.
        Then,
        the expected MSE in \eqref{eqn:emse_false},
        which is associated with the estimator $\Wbar=[\WbarS\T,\WbarF\T]\T$ in \eqref{eqn:lmmse_false} and $\WbarC=\bm 0$,
        can be decomposed as     
        \begin{align}\label{eqn:lemma:egen}
        \begin{split}
            \egen(p_S,p_C,p_F,n) & = 
            \egen_1 
             + \egen_2
            (\tr(\KxC) \! + \! \sigmav^2)
             + \egen_C,
        \end{split}
        \end{align}
        where 
        \begin{align}
            \egen_1 & = \! \eunder{\Abar}\big[
            \tr\!\big( 
                (\eye{p_S} \! \! - \! \WbarS \AS) \!
                \KxS
                (\eye{p_S} \! \! - \! \WbarS \AS)\!\T
            \big)\big], \label{eqn:lemma:egen_1} \\
            \egen_2 & = \eunder{\Abar}\big[
            \tr\!\big( \WbarS\WbarS\T \big)\big],
            \label{eqn:lemma:egen_2}\\
            \egen_C & = \tr(\KxC).
            \label{eqn:lemma:egen_C}
        \end{align}
        
        The expected MSE of the unknowns related to the fake features \eqref{eqn:egen_F_def} is
        \begin{align}\label{eqn:lemma:egen_F}\begin{split}
            \egen_F 
            & = \eunder{\Abar}\big[\tr\big( \WbarF \AS \KxS \AS\T \WbarF\T \big)\big] \\
            & \qquad + (\tr(\KxC) + \sigmav^2)\eunder{\Abar}\big[\tr\big( \WbarF\WbarF\T \big)\big]
        \end{split}
        \end{align}

    \end{lemma}
    
    \noindent Proof: See Appendix~\ref{proof:lemma:egen}.
    
        \begin{lemma}\label{lemma:Q:expectation}
            Let $\Amat\in\Rbb^{n \times p}$ be a standard Gaussian random matrix,
            $\Rmat = [\eye{p_S},\, \bm 0] \in \Rbb^{p_S\times p}$,
            $p > n$,
            $p>p_S$,
            $p_F = p - p_S$
            and $p>1$. 
            The matrix 
            $
                \Qmat = \Rmat\Amat\p\Amat\Rmat\T
                \Rmat\Amat\p\Amat\Rmat\T \in \Rbb^{p_S \times p_S}
            $
            has the expectation
            \begin{equation}
                \eunder{\Amat}\left[ \Qmat \right]
                = \left( 
                    \tfrac{n}{p}
                    - \tfrac{p_F n (p-n)}{(p-1)p(p+2)}
                \right)
                \eye{p_S}.
            \end{equation}
            Let $\RF = [\bm 0, \eye{p_F}]\in\Rbb^{p_F\times p}$.
            The matrix $\Qbar = \Rmat \Amat\p\Amat \RF\T \RF \Amat\p\Amat \Rmat\T\in \Rbb^{p_S\times p_S}$
            has the expectation
            \begin{equation}\label{eqn:lemma:E_Qbar}
                \eunder{\Amat}\left[ \Qbar \right]
                = \tfrac{np_F(p-n)}{(p-1)p(p+2)} \eye{p_S}.
            \end{equation}
            
        \end{lemma}
        
    \noindent Proof: See Appendix~\ref{proof:lemma:Q:expectation}.
    
    \begin{lemma}\label{lemma:haar_4th_moments}
        Let $\Vmat\in\Rbb^{p\times p}$ be a Haar distributed random orthogonal matrix \cite{meckes_2019},
        with its entries denoted by $v_{ij}$, 
        $i,\, j=1,\dots,p$,
        and $p>1$.
        Then the following moments hold
        if $i\neq j$, $k \neq l$,
        \begin{enumerate}
            \item $\eunder{\Vmat}[v_{il}^4] 
                        = \tfrac{3}{p(p+2)}$,
            \item $\eunder{\Vmat}[v_{il}^2 v_{ik}^2] 
                        = \eunder{\Vmat}[v_{il}^2 v_{jl}^2]
                        = \tfrac{1}{p(p+2)}$,
            \item $\eunder{\Vmat}[v_{il}v_{jl}v_{ik}v_{jk}]
                        = \tfrac{-1}{(p-1)p(p+2)} $.
        \end{enumerate}
        
    \end{lemma}
    
    \noindent Proof: See Appendix~\ref{proof:lemma:haar_4th_moments}.
    
    \begin{lemma}\label{lemma:AKAT}
        If $\Amat \in \Rbb^{n\times p}$ is a standard Gaussian matrix and $\Kmat\in\Rbb^{p\times p}$ a symmetric matrix,
        then
        $
            \eunder{\Amat}\left[\Amat\Kmat\Amat\T\right]
            = \tr(\Kmat) \eye{n}.
        $
        
    \end{lemma}
    
    \noindent Proof: See Appendix~\ref{proof:lemma:AKAT}.
    
    

\kern-0.25em
\subsection{Proof of Theorem~\ref{thm:emse_noisfree}}
\kern-0.25em
    \label{proof:thm:emse_noisfree}
    By Lemma~\ref{lemma:egen},
    we investigate $\egen_1$ and $\egen_2$ in \eqref{eqn:lemma:egen_1} and \eqref{eqn:lemma:egen_2}
    in the setting of Theorem~\ref{thm:emse_noisfree}.
    The estimator $\Wbar$ in the setting of Theorem~\ref{thm:emse_noisfree} is
    \begin{equation}\label{eqn:proof:Wbar_Abar_p}
        \Wbar 
        \!=\! \Khat_{\xbar} \Abar\T 
        (\Abar \Khat_{\xbar} \Abar\T \! + \! \Khat_{\vbar})\p 
        = \Abar\p = \begin{bmatrix}
            \WbarS \\ 
            \WbarF
        \end{bmatrix},
    \end{equation}
    where we have used that
    ${\Abar\p=\Abar\T(\Abar\Abar\T)\p\!}$.
    Recall that $\Abar=[\AS,\AF]$.
    With $\Rmat = [\eye{p_S},\bm 0]\in\Rbb^{p_S\times\pbar}$,
    we can write 
    \begin{equation}\label{eqn:proof:thm1:WbarS}
        \WbarS = \Rmat\Wbar = \Rmat\Abar\p,
    \end{equation}
    and $\AS=\Abar\Rmat\T$.
    We now combine $\egen_1$ and $\egen_2$ in \eqref{eqn:lemma:egen_1} and \eqref{eqn:lemma:egen_2},
    with \eqref{eqn:proof:thm1:WbarS},
    \begin{align}
        &\!\!\!\!\egen_1 \!  = \! \eunder{\Abar}\big[\!
        \tr\!\big( 
            (\eye{p_S} \! \! - \! \Rmat\Abar\p \!\!\Abar\Rmat\T)
            \KxS
            (\eye{p_S} \! \! - \! \Rmat\Abar\p \!\!\Abar\Rmat\T)\T
        \big)\big], \label{eqn:egen_1_RAPART}\\
        &\!\!\!\!\egen_2  
        \! = \! \eunder{\Abar}\big[\!
         \tr\!\big(
            \Rmat\Abar\p\Abar\pT\Rmat\T
        \big)
        = \eunder{\Abar}\big[\!
         \tr\!\big(
            \Rmat(\Abar\T\!\Abar)\p\Rmat\T
        \big)
        \big].
        \label{eqn:egen_2_RATAPRT}
    \end{align}
    
    We now analyze $\egen_1$ and $\egen_2$
    under the two scenarios of Theorem~\ref{thm:emse_noisfree},
    i.e., $n > \pbar + 1$ and $\pbar > n + 1$.

\kern 0.75em
\subsubsection{$n > \pbar + 1$}
    The matrix $\Abar \in \Rbb^{n \times \pbar}$ 
    is a standard Gaussian random matrix,
    hence if $n > \pbar + 1$,
    then $\Abar$ is full column rank with probability (w.p.) 1.
    It follows that $\Abar\p \Abar = \eye{\pbar}$ w.p. 1 if $n > \pbar + 1$.
    Hence, we obtain
    \begin{align}\label{eqn:egen1_n_geq_pbar}
        \egen_1 \! = \! \eunder{\Abar}\big[
        \tr\!\big( 
            (\eye{p_S} \! \! - \! \Rmat\Rmat\T) \!
            \KxS
            (\eye{p_S} \! \! - \! \Rmat\Rmat\T)\!
        \big)\big]
        = 0.
    \end{align}
    
    We note that the rows of $\Abar$,
    i.e., the columns of $\Abar\T$,
    are i.i.d. standard Gaussian vectors.
    From \cite[Prop. 1.2]{cook_forzani_wishart_2011},
    we have
    \begin{equation}\label{eqn:proof:inv_wishart}
        \eunder{\Abar}[(\Abar\T\Abar)\p] = \frac{1}{n-\pbar-1}\eye{\pbar}.
    \end{equation}
    Combining this expression with \eqref{eqn:egen_2_RATAPRT},
    we have
    \begin{equation}\label{eqn:egen_2_n_geq_pbar}
        \egen_2 = 
         \frac{1}{n-\pbar-1}
         \tr\!\big(
            \Rmat\Rmat\T
        \big)
        =  \frac{p_S}{n - \pbar - 1}.
    \end{equation}
    We now combine $ \egen_1$ from  \eqref{eqn:egen1_n_geq_pbar} and $ \egen_2$ from \eqref{eqn:egen_2_n_geq_pbar} 
    with \eqref{eqn:lemma:egen},
    to obtain \eqref{eqn:thm:emse_noisefree:n_geq_pbar} in Theorem~\ref{thm:emse_noisfree}.
    
\kern 0.75em
\subsubsection{$\pbar > n + 1$}
    For $\egen_2$,
    we note that now $\Abar\T\Abar\in\Rbb^{\pbar\times\pbar}$ is a singular matrix of rank $n$,
    and from \cite[Thm. 2.1]{cook_forzani_wishart_2011} we have
    \begin{equation}\label{eqn:proof:pinv_wishart}
        \eunder{\Abar}[(\Abar\T\Abar)\p] 
        = \frac{n}{\pbar(\pbar - n - 1)}\eye{\pbar},
    \end{equation}
    which combined with \eqref{eqn:egen_2_RATAPRT} gives
    \begin{equation}\label{eqn:egen_2_n_less_pbar}
        \egen_2 = \frac{n}{\pbar(\pbar - n - 1)}\tr(\Rmat\Rmat\T)
        = \frac{n p_S}{\pbar(\pbar - n - 1)}.
    \end{equation}
    
    We now expand the expression for $\egen_1$ in \eqref{eqn:egen_1_RAPART}, 
    and apply the cyclic property of the trace operator,
    \begin{align}\label{eqn:proof:egen_1_expanded}
    \begin{split}
        \egen_1 = \tr(\KxS)
        & + \eunder{\Abar}\big[\tr(\Rmat\Abar\p\Abar\Rmat\T\Rmat\Abar\p\Abar\Rmat\T\KxS) \\
        & - 2\tr(\Rmat\Abar\p\Abar\Rmat\T\KxS)\big].
    \end{split}
    \end{align}
    
    Let the full singular value decomposition (SVD) of $\Abar$ be denoted by
    $
        \Abar = \Umat \Smat \Vmat\T,
    $
    where $\Umat \in \Rbb^{n\times n}$,
    and $\Vmat \in \Rbb^{\pbar\times\pbar}$ 
    are Haar distributed orthogonal matrices 
    \cite[Section 2.1.5]{TulinoVerdu_2004},
    \cite{anderson_generation_1987},
    statistically independent from the diagonal matrix $\Smat\in\Rbb^{n\times \pbar}$,
    which contains the singular values of $\Abar$. 
    Denoting the columns of $\Vmat$ by $\vvec_i$,
    we have

    \begin{align}\label{eqn:expectation_Abarp_Abar}
        \eunder{\Abar}[\Abar\p\Abar]
        = \eunder{\Vmat\!,\,\Smat}[\Vmat\Smat\p\Smat\Vmat\T]
        = \!\!\!\!\sum_{i \in \mathcal{I}_{n/\pbar}} \eunder{\Vmat} [\vvec_i \vvec_i^T]
        = \frac{n}{\pbar} \eye{\pbar},
    \end{align}
    where $\mathcal{I}_{n/\pbar}$ is the set of $n$ indices out of $\{1, \ldots, \pbar\}$,  
    corresponding to non-zero singular values of $\Smat$.
    Here we have used that $\Ebb_{\Vmat}[v_{ij}^2]=1/\pbar$,
    and that $\Ebb_{\Vmat}[v_{ji}v_{li}]=0$ if $j\neq l$
    \cite[Section 2.1]{meckes_2019},
    where $v_{ij}$ denotes the entries of $\Vmat$.
    
    With 
    $
        \Qmat \! = \!
        \Rmat\Abar\p\Abar\Rmat\T\Rmat\Abar\p\Abar\Rmat\T
        \in\Rbb^{p_S\times p_S},
    $
    we now combine \eqref{eqn:expectation_Abarp_Abar}
    with \eqref{eqn:proof:egen_1_expanded},
    \begin{align}\label{eqn:proof:egen_1_trQmat}
        \egen_1 = \tr(\KxS)
        + \eunder{\Abar}\big[\tr(\Qmat\KxS)\big] 
        - 2\tfrac{n}{\pbar}\tr(\KxS).
    \end{align}
    From Lemma~\ref{lemma:Q:expectation}
    we have 
    that $\Ebb_{\Abar}[\Qmat] = \mu_q\eye{p_S}$,
    with
    $
        \mu_q = \frac{n}{\pbar}
            - \frac{p_Fn(\pbar-n)}{(\pbar-1)\pbar(\pbar+2)}
    $
    which we now apply to \eqref{eqn:proof:egen_1_trQmat}.
    Hence,
    \begin{align}
        \egen_1 
        &= \big(
            1 -\tfrac{n}{\pbar} -\tfrac{p_F n (\pbar - n)}{(\pbar - 1)\pbar(\pbar + 2)}
        \big)
        \tr(\KxS),
        \label{eqn:egen_1_n_less_pbar}
    \end{align}
    Combining \eqref{eqn:egen_1_n_less_pbar} and 
    \eqref{eqn:egen_2_n_less_pbar} with \eqref{eqn:lemma:egen},
    we find the desired expression of $\egen$ in \eqref{eqn:thm:emse_noisefree:n_less_pbar} in Theorem~\ref{thm:emse_noisfree}.


\kern-0.25em
\subsection{Proof of Corollary~\ref{lemma:local_minima_overp}}
\kern-0.5em
    \label{proof:lemma:local_minima_overp}
    We first prove \textit{i)} of Corollary~\ref{lemma:local_minima_overp}.
    Let $p_F\rightarrow\infty$,
    hence $n<\pbar=p_S+p_F$.
    Now consider the expression for $\egen$ in \eqref{eqn:thm:emse_noisefree:n_less_pbar} of Theorem~\ref{thm:emse_noisfree}.
    Furthermore,
    with $n$ and $p_S$ constant and finite,
    we have that the expression in front of $\tr(\KxS)$ goes to one, 
    and the fraction in front of $\tr(\KxC) + \sigmav^2$ goes to zero.
    Hence,
    $\egen\rightarrow \tr(\KxS) + \tr(\KxC) = \tr(\Kx)$.
    
    We now prove \textit{ii)}.
    The derivative of $\egen$ in 
    \eqref{eqn:thm:emse_noisefree:n_less_pbar}
    w.r.t. $p_F$ is 
    \begin{align}\begin{split}
        \tfrac{\partial \egen}{\partial p_F}
            & = -\tfrac{np_S(2\pbar -n -1)}{\pbar^2(\pbar-n-1)^2}(\tr(\KxC) + \sigmav^2)\\
            & ~\quad + \Big( 0 + \tfrac{n}{\pbar^2}
            - \tfrac{(n(\pbar-n) + p_F n )(\pbar-1)\pbar(\pbar+2)}{(\pbar-1)^2\pbar^2(\pbar+2)^2} \\
            & \quad\qquad + \tfrac{p_Fn(\pbar-n)(3\pbar^2+2\pbar-2)}{(\pbar-1)^2\pbar^2(\pbar+2)^2}
            \Big) \tr(\KxS).
            \label{eqn:proof:lemma:d_egen_d_pf}
        \end{split}
    \end{align}
    We now let $p_F\rightarrow\infty$,
    and analyze the proportionality of the expression in \eqref{eqn:proof:lemma:d_egen_d_pf} with respect to (w.r.t.) $p_F$:
    \begin{align*}
        \begin{split}
            \tfrac{\partial \egen}{\partial p_F}
            \propto & - \tfrac{2 n p_S p_F}{p_F^4}
            (\tr(\KxC) + \sigmahatvbar^2) + \Big(\tfrac{2n}{p_F^2} \\
            & - \tfrac{2n(n+p_S+1)p_F}{p_F^4} + \tfrac{3n(p_S-1)(n-1)p_F^2}{p_F^6}\Big)\tr(\KxS)
        \end{split} \\
        \propto & - \tfrac{2 n p_S}{p_F^3}
            (\tr(\KxC) \! + \! \sigmahatvbar^2)
            \! + \! \tfrac{2n}{p_F^2}\tr(\KxS)
        \propto \tfrac{2n}{p_F^2}\tr(\KxS).
    \end{align*}
    Hence we have shown that if $p_F\rightarrow\infty$,
    then the derivative of $\egen$ w.r.t. $p_F$ approaches zero from the positive side.
    In other words,
    the expected MSE approaches its limit of $\egen\rightarrow\tr(\Kx)$ from below. 
    This concludes the proof.
\kern-0.25em
\subsection{Proof of Corollary~\ref{lemma:power_ratio_limits}}
\kern-0.5em
    \label{proof:lemma:power_ratio_limits}
    From Corollary~\ref{lemma:local_minima_overp},
    we have that if $p_F\rightarrow\infty$
    and $n$, $p_S$ and $p_C$ are finite,
    then $\egen \rightarrow \egen_\infty \triangleq \tr(\Kx)$.
    
    If $n \! > \! p_S \! + \! 1$ and $p_F\! =\! 0$,
    then $\egen = \egen_0 \triangleq \tr(\KxC) + \frac{p_S}{n-p_S-1}(\tr(\KxC) + \sigmav^2)$.
    Inserting $\tr(\KxS) = r\tr(\Kx)$
    and $\tr(\KxC) = (1-r)\tr(\Kx)$ 
    into the inequality 
    $\egen_\infty < \egen_0$ 
    and solving for $r$ leads to the expression in \eqref{eqn:r_bound_n_geq_ps}.


If $p_S\!>\!n+1$ and $p_F\!=\!0$,
then we have 
$\egen_0 \! = \tr(\KxC) + \! \frac{n}{p_S-n-1}(\tr(\KxC) + \sigmav^2) + (1-\frac{n}{p_S})\tr(\KxS)$.
Again, inserting $\tr(\KxS) \!  =  \! r\tr(\Kx)$ and $\tr(\KxC) \!  = \!  (1 \! - \! r)\tr(\Kx)$ into $\egen_\infty<\egen_0$,
and solving for $r$ gives the desired expression in \eqref{eqn:r_bound_n_leq_ps}.

\kern-0.25em
\subsection{Proof of Theorem~\ref{thm:emse_noise_model}}
\kern-0.5em
    \label{proof:thm:emse_noise_model}
    By Lemma~\ref{lemma:egen},
    we investigate $\egen_1$ and $\egen_2$ in \eqref{eqn:lemma:egen_1} and \eqref{eqn:lemma:egen_2},
    in the setting of Theorem~\ref{thm:emse_noise_model}.
    The estimator in this setting is 
    \begin{equation}\label{eqn:proof:wbar:thm2}
        \Wbar = \Abar\T(\Abar\Abar\T + \sigmahatvbar^2\eye{n})\inv.
    \end{equation}
    We first investigate the term $\egen_1$ from \eqref{eqn:lemma:egen_1}.
    Throughout the proof, 
    we use the following notation for the full SVD of $\Abar$:
    \begin{equation}\label{eqn:proof:svd_A}
        \Abar = \Umat \Smat \Vmat\T,
    \end{equation}
    where $\Umat \in \Rbb^{n\times n}$,
    and $\Vmat \in \Rbb^{\pbar\times\pbar}$ 
    are Haar distributed orthogonal random matrices \cite[Section 2.1.5]{TulinoVerdu_2004},
    \cite{anderson_generation_1987},
    statistically independent from $\Smat\in\Rbb^{n\times \pbar}$,
    which contains the singular values $s_i$ of $\Abar$,
    $i=1,\dots,\min\{n,\pbar\}$,
    with $s_i = 0$ if $i>\min\{n,\pbar\}$.
    With $\Rmat = [\eye{p_S}, 0]\in\Rbb^{p_S\times \pbar}$,
    we have $\WbarS=\Rmat\Wbar$.
    Now let $\Mmat\in\Rbb^{p_S\times p_S}$ be defined by
    \begin{align}
        \Mmat & = \eye{p_S} \! - \! \WbarS\AS 
        = \eye{p_S} \! - \! \Rmat\Abar\T\!(\Abar\Abar\T \! + \! \sigmahatvbar^2\eye{n})\inv \! \Abar\Rmat\T \nonumber
        \\
        &= \eye{p_S} - \Rmat \Vmat\Smat\T(\Smat\Smat\T + \sigmahatvbar^2\eye{n})\inv\Smat\Vmat\T\Rmat\T \\
        &= \Rmat\Vmat\Big(\eye{\pbar
        } - 
        \diag\Big(\tfrac{\lambda_i}{\lambda_i+\sigmahatvbar^2}\Big)
        \Big)\Vmat\T\Rmat\T\\
        &= \Rmat\Vmat
        \diag\big(\Tilde{\lambda}_i\big)
        \Vmat\T\Rmat\T,
    \end{align}
    where $\Tilde{\lambda}_i = \sigmahatvbar^2/(\lambda_i + \sigmahatvbar^2)$,
    and $\lambda_i=s_i^2$ denote the eigenvalues of $\Abar\T\Abar\in\Rbb^{\pbar\times\pbar}$.
    Inserting this into $\egen_1$ in \eqref{eqn:lemma:egen_1},
    we have
    \begin{align}\label{eqn:proof:egen_1:tr_E_Qmat_Kxs}
        \egen_1 &= 
        \eunder{\Amat}\Big[\tr\!\big( 
            \Mmat
            \KxS
            \Mmat\T
        \big)\Big] 
        =\tr\!\left( 
             \eunder{\Amat}
             \left[\Qmat\right]
            \KxS
        \right),
    \end{align}
    where $\Qmat = \Mmat\T\Mmat\in\Rbb^{p_S\times p_S}$ has the entries
    $q_{ij}=\sum_{k=1}^{p_S} m_{ki}m_{kj}$,
    where $m_{ij} = \sum_{l=1}^{\pbar} v_{il}v_{jl} \Tilde{\lambda}_l$ denotes the $(i,j)$\th{} entry of $\Mmat$.
    
    We now investigate the diagonal and off-diagonal entries of 
    $\Qmat$ in expectation.
    The diagonal entries are
    $
        q_{ii}= \sum_{k=1}^{p_S} m_{ki}^2,
    $
    which has one term where $k=i$:
    \begin{equation*}
        m_{ii}^2 = \Big(\sum_{l=1}^{\pbar} v_{il}^2 \Tilde{\lambda}_l \Big)^2
        = \sum_{l=1}^{\pbar} \Big(v_{il}^4\Tilde{\lambda}_l^2 + 2\sum_{j=1}^{l-1}v_{il}^2v_{ij}^2 \Tilde{\lambda}_l\Tilde{\lambda}_j\Big),
    \end{equation*}
    and $(p_S-1)$ terms where $k\neq i$:
    \begin{align*}
        m_{ki}^2 
        & = \sum_{l=1}^{\pbar} \Big(v_{kl}^2v_{il}^2 \Tilde{\lambda}_l^2 
        + 2\sum_{j=1}^{l-1} v_{kl}v_{il}v_{kj}v_{ij} \Tilde{\lambda}_l\Tilde{\lambda}_j
        \Big).
    \end{align*}
    Using that the random matrix $\Vmat$ is uncorrelated with $\Smat$,
    and the expectations from Lemma~\ref{lemma:haar_4th_moments},
    we have the expectations
    \begin{equation*}
        \eunder{\Vmat\!\!,\Smat}[m_{ii}^2] 
        = \sum_{l=1}^{\pbar} \Big(\tfrac{3}{\pbar(\pbar+2)}  \eunder{\Smat}[\Tilde{\lambda}_l^2]
        + 2\sum_{j=1}^{l-1}\tfrac{1}{\pbar(\pbar+2)} \eunder{\Smat}[\Tilde{\lambda}_l\Tilde{\lambda}_j]\Big),
    \end{equation*}
    \begin{align*}\begin{split}
        \eunder{\Vmat\!\!,\Smat}[m_{ki}^2]
        &= \sum_{l=1}^{\pbar}\Big(
            \tfrac{1}{\pbar(\pbar+2)}\eunder{\Smat}[\Tilde{\lambda}_l^2] 
        - 2\sum_{j=1}^{l-1}\tfrac{1}{(\pbar-1)\pbar(\pbar+2)}\eunder{\Smat}[\Tilde{\lambda}_l\Tilde{\lambda}_j]
        \Big),
    \end{split}
    \end{align*}
    which together with $q_{ii}= \sum_{k=1}^{p_S} m_{ki}^2$,
    gives 
    \begin{align}\label{eqn:proof:E_q_ii}
        \eunder{\Amat}[q_{ii}] = \muM_2,
    \end{align}
    with $\muM_2$ as in \eqref{eqn:thm:muM_2}.
    
    The off-diagonal entries of $\Qmat=\Mmat\T\Mmat$ are,
    with $i\neq j$,
    \begin{align*}
        q_{ij}
        = \sum_{k=1}^{p_S} m_{ki}m_{kj}
        = \sum_{k=1}^{p_S} \bigg(
                \sum_{l=1}^{\pbar} v_{kl}v_{il}\Tilde{\lambda}_l
            \bigg)\bigg(
                \sum_{l=1}^{\pbar} v_{kl}v_{jl}\Tilde{\lambda}_l
            \bigg).
    \end{align*}
    By \cite[Lemma 2.22]{meckes_2019},
    products of entries from $\Vmat$ are zero-mean if any row- or column-index occurs an odd number of times in the product.
    Hence
    \begin{equation}\label{eqn:proof:E_q_ij}
        \eunder{\Amat}[q_{ij}] = 0, ~ i\neq j,
    \end{equation}
    since in each term of $q_{ij}$, 
    there is an odd number of entries from row $i$ and $j$ of $\Vmat$.
    By \eqref{eqn:proof:E_q_ii} and \eqref{eqn:proof:E_q_ij},
    we have now that $\Ebb_{\Amat}[\Qmat] = \Bar{\mu}_2 \eye{p_S}$,
    and together with \eqref{eqn:proof:egen_1:tr_E_Qmat_Kxs}
    we find that 
    \begin{equation}\label{eqn:proof_sigmahat_final_e1}
        \egen_1 = \Bar{\mu}_2 \tr(\KxS).
    \end{equation}
    
    We now find $\egen_2$ from \eqref{eqn:lemma:egen_2}
    in the setting of Theorem~\ref{thm:emse_noise_model}.
    Using the SVD of $\Abar$ in \eqref{eqn:proof:svd_A},
    and that $\WbarS=\Rmat\Wbar$, and \eqref{eqn:proof:wbar:thm2},
    we write
    \begin{align}\label{eqn:proof:WbarSWbarST}
        \WbarS\WbarS\T 
        & \! = \! \Rmat\Wbar\Wbar\T\Rmat\T 
        \! =\! \Rmat\Vmat\Smat\T
        (\Smat\Smat\T\!\!+\!\!\sigmahatvbar^2\eye{n})^{-2}
        \Smat\Vmat\T\Rmat\T
        \nonumber
        \\
        & = \!\Rmat\Vmat\diag\Big(\tfrac{\lambda_i}{(\lambda_i+\sigmahatvbar^2)^2}\Big)
        \Vmat\T\Rmat\T.
    \end{align}
    We combine \eqref{eqn:proof:WbarSWbarST} with \eqref{eqn:lemma:egen_2},
    and use that $\Vmat$ and $\Smat$ are statistically independent,
    \begin{align}\label{eqn:proof:egen_2:VTRTRV}
        \egen_2 
        &=  
        \tr\Big(\eunder{\Smat}\Big[ 
                \diag\Big(
                    \tfrac{\lambda_i}{(\lambda_i+\sigmahatvbar^2)^2}
                \Big)
            \Big]
            \eunder{\Vmat}\big[ 
                \Vmat\T\Rmat\T\Rmat\Vmat
            \big]
        \Big).
    \end{align}
    With $\vvec_i\T\in\Rbb^{1\times\pbar}$ denoting the rows of $\Vmat$,
    we have that
    $\eunder{\Vmat}[\Vmat\T\Rmat\T\Rmat\Vmat] = \sum_{i=1}^{p_S}\Ebb_{\Vmat}[\vvec_i\vvec_i\T]=p_S\frac{1}{\pbar}\eye{\pbar}$.
    Here, we have used that $\Ebb_{\Vmat}[v_{ij}^2]=\frac{1}{\pbar}$,
    and that $\Ebb_{\Vmat}[v_{ij}v_{il}]=0$ if $j\neq l$ \cite[Section 2.1]{meckes_2019}.
    Combining this with \eqref{eqn:proof:egen_2:VTRTRV},
    we find
    \begin{equation}\label{eqn:proof:egen_2_final}
       \textstyle \egen_2 
        =  \frac{p_S}{\pbar}
        \sum_{i=1}^{\pbar}\eunder{\lambda_i}\Big[ 
              \frac{\lambda_i}{(\lambda_i+\sigmahatvbar^2)^2}
            \Big].
    \end{equation}
    
    We now combine \eqref{eqn:proof_sigmahat_final_e1} 
    and \eqref{eqn:proof:egen_2_final},
    with \eqref{eqn:lemma:egen} and
    find the desired expression of $\egen$ 
    in \eqref{eqn:thm2:emse} of Theorem~\ref{thm:emse_noise_model}.

\kern-0.25em
\subsection{Proof of Corollary~\ref{col:n_gg_pbar}}\label{proof:col:n_gg_pbar}
\kern-0.25em
    For $n\gg\pbar$,
    by the law of large numbers,
    $\Abar\T\Abar\approx \Ebb[\Abar\T\Abar] = n\eye{\pbar}$,
    and $\lambda_i\approx n$, $i=1,\dots,\pbar$.
    Substituting $\lambda_i\approx n$ into the eigenvalue expressions for $\muM_1$ and $\muM_2$ in \eqref{eqn:thm:muM_1} and \eqref{eqn:thm:muM_2} of Theorem~\ref{thm:emse_noise_model},
    we obtain the expressions in \eqref{eqn:muM_1_and_2_n_gg_pbar},
    and the desired approximation for $\egen$ in \eqref{eqn:approx_n_gg_pbar}.

\kern-0.25em
\subsection{Proof of Lemma~\ref{lem:optimal_sigmahat}}
\kern-0.25em
    \label{proof:lem:optimal_sigmahat}
    For $p_F=0$,  we have $\pbar = p_S$.
    With $r_{\min} = \min(n,p_S)$,
    $\muM_2$ of Theorem~\ref{thm:emse_noise_model} is given by 
    $
        \muM_2 = 
    \frac{r_{\min}}{p_S}\eunder{\lambda}[
        \frac{\sigmahatvbar^4}{(\lambda + \sigmahatvbar^2)^2}
    ] + \frac{p_S-r_{\min}}{p_S}.
    $
    Taking partial derivative with respect to $\sigmahatvbar^2$, we obtain
    $
        \frac{\partial \muM_2}{\partial \sigmahatvbar^2}
        = \frac{2\sigmahatvbar^2r_{\min}}{p_S}
        \Ebb[\frac{\lambda}{(\lambda+\sigmahatvbar^2)^3}].
    $
   Similarly, 
    $
        \frac{\partial \muM_1}{\partial \sigmahatvbar^2}
        = -2 r_{\min} \Ebb[\frac{\lambda}{(\lambda+\sigmahatvbar^2)^3}].
    $
    Hence, 
        $
        \tfrac{\partial \egen}{\partial \sigmahatvbar^2}
        = (\tfrac{2\sigmahatvbar^2}{p_S}\tr(\KxS) - 2 (\tr(\KxC) + \sigmav^2))r_{\min}
        \Ebb[\tfrac{\lambda}{(\lambda + \sigmahatvbar^2)^3}],
        $
    where the expectation term is always non-negative. 
    Setting the derivative 
    $\frac{\partial \egen}{\partial \sigmahatvbar^2}$ 
    to zero, 
    we find the optimal $\sigmahatvbar^2$ as in \eqref{eq:optimal_sigmahat}. 
    Note that $\frac{\partial^2\egen}{\partial (\sigmahatvbar^2)^2} > 0$,
    hence we indeed find a minimum of the function.
    We note that for the other stationary point at $\sigmahatvbar\rightarrow\infty$,
     $\frac{\partial^2\egen}{\partial (\sigmahatvbar^2)^2} < 0$,
    hence this point is not a minimum.
    

    For $n\gg\pbar$, $p_F>0$,
    we have the approximation for $\egen$ in \eqref{eqn:approx_n_gg_pbar} from Corollary~\ref{col:n_gg_pbar}. 
    We take the derivative of this approximation w.r.t. $\sigmahatvbar^2$,
    and find the same solution for $\sigmahatvbar$.

\kern-0.25em
\subsection{Proof of Theorem~\ref{thm:egen_F}}\label{sec:proof:thm:egen_F}
\kern-0.5em
    The line of argument is similar to the proof of Theorem~\ref{thm:emse_noisfree}, hence here we only present the key steps. Let $\RF = [\bm 0, \eye{p_F}]\in\Rbb^{p_F\times \pbar}$.
    By \eqref{eqn:proof:Wbar_Abar_p}, we have that $\WbarF = \RF \Abar\p$,
    and  
    $\Ebb_{\Abar}[\WbarF\WbarF\T] = \RF \Ebb_{\Abar}[(\Abar\T\Abar)\p] \RF\T $.
    Note that $\RF\RF\T = \eye{p_F}$. Hence, 
    if $n>\pbar + 1$,
    then $\tr(\Ebb_{\Abar}[\WbarF\WbarF\T]) = \frac{p_F}{n - \pbar - 1}$, similar to \eqref{eqn:proof:inv_wishart}.  
    If instead $\pbar > n + 1$,
    then $\tr(\Ebb_{\Abar}[\WbarF\WbarF\T]) = \frac{p_Fn}{\pbar(\pbar- n - 1)}$, similar to \eqref{eqn:egen_2_n_less_pbar}. 
    Hence, we have derived the second term of \eqref{eqn:lemma:egen_F}.
    We now consider the first term of \eqref{eqn:lemma:egen_F}.
    If $n > \pbar + 1$,
    then $\Abar\p\Abar = \eye{\pbar}$, w.p. 1,
    hence $\WbarF\AS = \RF\Abar\p\Abar\Rmat\T = [\bm 0, \eye{p_F}]\eye{\pbar}[\eye{p_S}, \bm 0]\T = \bm 0$ with $\Rmat = [\eye{p_S}, 0]\in\Rbb^{p_S\times \pbar}$, 
    hence the first term of \eqref{eqn:lemma:egen_F} is zero.
    If instead $\pbar > n + 1$,
    we note that $\AS\T\WbarF\T\WbarF\AS = \Rmat \Abar\T\Abar\pT \RF\T\RF \Abar\p\Abar \Rmat$
    and apply \eqref{eqn:lemma:E_Qbar} to obtain
    $
        \eunder{\Abar}[\tr(\AS\T\WbarF\T\WbarF\AS\KxS)] 
        = \frac{n p_F(\pbar - n)}{(\pbar - 1)\pbar(\pbar+2)} \tr(\KxS),
    $
    which concludes the proof.

\kern-0.5em
\subsection{Proof of Lemma~\ref{lemma:egen}}
\kern-0.75em
    \label{proof:lemma:egen}
    By \eqref{eqn:lmmse_false}, $\Wbar$ is given by
     $\Wbar=\Khatxbar\Abar\T(\Abar\Khatxbar\Abar\T+\Khatvbar)\p = [\WbarS\T,\WbarF\T]\T$.
     Note that $\Wbar$ depends on $\AS$ and $\AF$, but not on $\AC$.
    Combining \eqref{eqn:mse_WbarC_zero}
    with \eqref{eqn:emse_false},
    \begin{equation}\label{eqn:lemma:proof:egen_decomp}
        \egen
        = \Ebb_{\Amat}[J_S(\WbarS)] + \Ebb_{\Amat}[J_C(\WbarC)]
        = \egen_S + \egen_C,
    \end{equation}
    where 
    $\egen_S=\Ebb_{\Amat}[J_S(\WbarS)]$ and
    $\egen_C = \Ebb_{\Amat}[J_C(\WbarC)]$.
    Here,
    we have dropped the arguments of $\egen(p_S,p_C,p_F,n)$ for ease of disposition,
    and we recall that $\Amat = [\AS,\AC,\AF]$.
    %
    
    We now investigate $\egen_C$ in \eqref{eqn:lemma:proof:egen_decomp}. 
    Using $\WbarC=\bm 0$, 
    i.e., $\xChat = \WbarC\yvec = \bm 0$, we have
    \begin{equation}
        J_C(\WbarC) 
        = \Ebb_{\xvec,\yvec}[\|\xC - \xChat\|^2] 
        = \tr(\KxC).
    \end{equation}
    We obtain  $\egen_C= \Ebb_{\Amat}[J_C(\WbarC) ] = \tr(\KxC)$,
    which
    matches the desired expression in \eqref{eqn:lemma:egen_C}.
    
    We now investigate the term $J_S(\WbarS)$,
    as defined by \eqref{eqn:JS_WS}.
    With $\xShat= \WbarS\yvec$,
    and $\yvec$ from the underlying system in \eqref{eqn:model_true},
    \begin{align}\label{eqn:MSE_trace_xhat_S}
        & \!\!\! J_S(\WbarS) 
        \! = \eunder{\xvec, \vvec}\big[
                \|\xS \! - \! \WbarS(\AS\xS + \AC\xC + \vvec)\|^2
            \big] \\
        \begin{split}
            & \!\!\!  = \eunder{\xS}[
                    \|(\eye{p_S} \! \! - \! \WbarS \AS) \xS\|^2
                ] 
                + \eunder{\xC}[
                    \|\WbarS \AC\xC\|^2
                ] \\
            & \!\!\! \quad + \eunder{\vvec}[
                    \|\WbarS\vvec\|^2
                ] 
            \! - \! 2\eunder{\xvec}[
                    \xS\T(\eye{p_S} \! \! - \! \WbarS \AS)\T
                    \WbarS\AC\xC
                ],
        \end{split}
    \end{align}
    where we have used that $\vvec$ is zero-mean and statistically independent from $\xS$ and $\xC$,
    and eliminated the associated cross-terms.
    Rewriting with the trace operator,
    \begin{align}
    \begin{split}\label{eqn:trace_MSE_wbarS_1}
    J_S(\WbarS) \!
        &=   \tr\!\big(
                (\eye{p_S} \! \! - \! \WbarS \AS) \!
                \KxS
                (\eye{p_S} \! \! - \! \WbarS \AS)\!\T
            \big) \\
         & \! + \! \tr\!\big(
                \WbarS\AC
                \KxC
                \AC\T\WbarS\T
            \big)
        \!+\! \tr\!\big(
                \WbarS 
                \Kv
                \WbarS\T
            \big) \\
        & \! - 2\tr\!\big(
                \WbarS\AC
                \KxCxS
                (\eye{p_S} \! \! - \! \WbarS \AS)\!\T
            \big),
    \end{split}
    \end{align}
    where we have used that $\KxS = \Ebb_{\xS}[\xS\xS\T]$,
    $\KxC=\Ebb_{\xC}[\xC\xC\T]$,
    $\Kv = \Ebb_{\vvec}[\vvec\vvec\T]$ and 
    $\KxCxS=\Ebb_{\xvec}[\xC\xS\T]$.
    
    We now consider \eqref{eqn:trace_MSE_wbarS_1} in expectation over the regressor matrices, i.e., $\egen_S=\Ebb_{\Amat}[J_S(\WbarS)]$.
    We eliminate the cross-terms between $\AC$ and $\Wbar$,
    and between $\AC$ and $\AS$,
    due to statistical independence,
    \begin{align}
    \label{eqn:es:Ws}
    \begin{split}
        \egen_S
        & = \eunder{\Abar}\big[
        \tr\!\big( 
                (\eye{p_S} \! \! - \! \WbarS \AS) \!
                \KxS
                (\eye{p_S} \! \! - \! \WbarS \AS)\!\T
            \big) \\
        & \quad + \tr\!\big(
                    \WbarS
                    \big(
                        \eunder{\AC}\big[\AC\KxC\AC\T\big]
                        + \Kv
                    \big)
                    \WbarS\T
                \big)
        \big]
    \end{split}
    \\
    & = \egen_1 + \egen_2,
    \label{eqn:egen_before_n_geq_pbar}
    \end{align}
    with $\egen_1$ and $\egen_2$ 
    as in the desired expressions \eqref{eqn:lemma:egen_1} and \eqref{eqn:lemma:egen_2}.
    Note that in the final step,
    we used that 
    $\Kv = \sigmav^2\eye{n}$,
    and that
    $\Ebb_{\AC}\left[\AC\KxC\AC\T\right] = \tr(\KxC) \eye{n}$,
    from Lemma~\ref{lemma:AKAT},
    which we can apply due to $\AC$ being standard Gaussian.

    Using that $\xF=\bm 0$,
    we have $\|\xF-\xFhat\|^2 = \|\xFhat\|^2$ and
    \begin{align}
        J_F(\WbarF) 
        & = \eunder{\xvec,\vvec}[\|\WbarF(\AS\xS+\AC\xC+\vvec)\|^2],
    \end{align}
    and by using similar steps as for $J_S$,
    we expand the norm,
    cancel the cross-terms,
    rewrite the expression with the trace operator,
    take the expectation over $\Amat$, 
    and find $\egen_F$ of Lemma~\ref{lemma:egen}.

\kern-0.6em
\subsection{Proof of Lemma~\ref{lemma:Q:expectation}}
\kern-0.5em
\label{proof:lemma:Q:expectation}
    Let the full SVD of $\Amat$ be denoted by
    $
        \Amat = \Umat \Smat \Vmat\T,
    $
    where $\Umat\in\Rbb^{n\times n}$ and $\Vmat\in\Rbb^{p\times p}$
    are Haar-distributed orthogonal random matrices
    statistically independent from 
    and $\Smat\in\Rbb^{n\times p}$ 
    \cite[Section 2.1.5]{TulinoVerdu_2004},
    \cite{anderson_generation_1987},
    which contains the singular values of $\Amat$.
    %
    Letting $\vvec_i$ denote the columns of $\Vmat$,
    we have
    \begin{align}
        \Mmat & = \textstyle \Vmat\Smat\p\Smat\Vmat\T 
        =\sum_{i\in\Ic_{n/p}}\vvec_i\vvec_i\T,
    \end{align}
    where $\mathcal{I}_{n/p}$ is the set of $n$ indices out of $\{1, \ldots, p\}$,  
    corresponding to non-zero singular values of $\Smat$.
    Noting that $\Smat$ is of rank $n$ w.p. 1,
    we choose $\mathcal{I}_{n/p} = \{1,\dots,n\}$,
    without loss of generality 
    due  to 
    the Haar distribution of $\Vmat$. In other words, 
    we henceforth write the $(i,j)$\th{} entry of $\Mmat$ as
    $
        m_{ij} = \sum_{k=1}^n v_{ik}v_{jk}.
    $
    
    The matrix $\Qmat\in\Rbb^{p_S\times p_S}$ can be written as 
    $
        \Qmat = \Rmat\Mmat\Rmat\T
                \Rmat\Mmat\Rmat\T,
    $
    where $\Rmat=[\eye{p_S}, \bm  0]\in\Rbb^{p_S\times p}$,
    hence the $(i,j)$\th{} entry of $\Qmat$
    is
    $
        q_{ij} = \sum_{l=1}^{p_S} m_{il}m_{lj} 
        =\sum_{l=1}^{p_S} m_{il}m_{jl},
    $
    where we have used $\Mmat\T=\Mmat$.
    The diagonal elements $q_{ii}$ can be written as
    $q_{ii} \!=\! \sum_{l=1}^{p_S} m_{il}^2
        = m_{ii}^2 
        \!+\! \sum_{{j=1,\,j\neq i}}^{p_S} m_{ij}^2$,
    which we further expand,
    \begin{align}
    \begin{split}
        q_{ii}
        & = 
            \sum_{l=1}^n 
                v_{il}^4 
            + 2 \sum_{l=1}^n\sum_{k=1}^{l-1} 
                v_{il}^2v_{ik}^2\\
        & \quad 
        + \sum_{{j=1,\,j\neq i}}^{p_S}\Big(
            \sum_{l=1}^n 
                v_{il}^2v_{jl}^2 
            + 2\sum_{l=1}^n\sum_{k=1}^{l-1}
                v_{il}v_{jl}v_{ik}v_{jk}\Big).
    \end{split}
    \end{align}
    
    Lemma \ref{lemma:haar_4th_moments} gives the moments necessary to derive the expectation of $q_{ii}$.
    Combining these moments, 
    we find
    \begin{align}
        \mu_q \triangleq \eunder{\Amat}[q_{ii}]
        & = \tfrac{n}{p(p+2)}\left( n+p_S+1 - \tfrac{(p_S-1)(n-1)}{p-1}\right).
    \label{eqn:proof:lemma:mu_q}
    \end{align}
    %
    Regarding the off-diagonal entries of $\Qmat$,
    we have with $i\neq j$
    \begin{align}\label{eqn:proof:thm2:q_ij}
       \textstyle q_{ij} = 
        m_{ii}m_{ij} + m_{ij}m_{jj} 
        + \sum_{{l=1,\,l\neq i,j}}^{p_S} m_{il}m_{lj}.
    \end{align}
    By \cite[Lemma 2.22]{meckes_2019},
    products of entries from $\Vmat$ are zero-mean if any row- or column-index occurs an odd number of times in the product.
    Hence
    $
        m_{ii}m_{ij} = (v_{i1}^2+\cdots+v_{in}^2)(v_{i1}v_{j1}+\cdots+v_{in}v_{jn}),
    $
    is zero-mean, due to the row-index $i$ occurring three times in each term of the summation.
    Similarly, the other terms of \eqref{eqn:proof:thm2:q_ij},
    are also zero-mean.
    Hence $\Ebb_{\Amat}[q_{ij}]=0$ for $i\neq j$.
    Combining this with 
    \eqref{eqn:proof:lemma:mu_q}, we obtain the sought for expressions of $\Ebb_{\Amat}[\Qmat]$.
    
    The diagonal elements of $\Qbar\in\Rbb^{p_S\times p_S}$ can be written as
    \begin{equation}
        \qbar_{ii} = \sum_{l=p_S+1}^p \sum_{k=1}^n
        \Big(
            v_{ik}^2 v_{lk}^2 + 2\sum_{j=1}^{k-1} v_{ik}v_{lk}v_{ij}v_{lj},
        \Big),
    \end{equation}
    which in expectation is, using Lemma~\ref{lemma:haar_4th_moments},
    \begin{equation}\label{eqn:muqbar}
        \mu_{\qbar} \triangleq \eunder{\Amat}[\qbar_{ii}] = p_F n \Big(
            \tfrac{1}{p(p+2)} - \tfrac{n-1}{(p-1)p(p+2)}
        \Big).
    \end{equation}
    Similarly as for $q_{ij}$ in \eqref{eqn:proof:thm2:q_ij},
    which we have shown is zero-mean over the distribution of $\Amat$,
    the off-diagonal entries in $\Qbar$ are also zero-mean.
    Combining this with \eqref{eqn:muqbar}
    and simplifying,
    we obtain the sought for expression of $\Ebb_{\Amat}[\Qbar]$.

\kern-0.5em
\subsection{Proof of Lemma~\ref{lemma:haar_4th_moments}}
\kern-0.125em
\label{proof:lemma:haar_4th_moments}
    Let $\delta_{\alpha\beta}=1$ if $\alpha=\beta$, and zero otherwise. 
    Applying \cite[Lemma 2.22]{meckes_2019}, we obtain  
    \begin{align*}
       \eunder{\Vmat}[v_{il}^4] 
        & = \tfrac{-1}{(p-1)p(p+2)}\big[
            \delta_{ii}\delta_{ii}\delta_{ll}\delta_{ll}
            +\delta_{ii}\delta_{ii}\delta_{ll}\delta_{ll}
            +\delta_{ii}\delta_{ii}\delta_{ll}\delta_{ll} \\
        & \quad\qquad
            +\delta_{ii}\delta_{ii}\delta_{ll}\delta_{ll}
            +\delta_{ii}\delta_{ii}\delta_{ll}\delta_{ll}
            +\delta_{ii}\delta_{ii}\delta_{ll}\delta_{ll}
        \big] \\
        &\quad + \tfrac{p+1}{(p-1)p(p+2)}\big[
            \delta_{ii}\delta_{ii}\delta_{ll}\delta_{ll}
            +\delta_{ii}\delta_{ii}\delta_{ll}\delta_{ll}
            +\delta_{ii}\delta_{ii}\delta_{ll}\delta_{ll}
        \big] \\
        & = \tfrac{-1\left[
            1 +1 +1 +1 +1 +1
        \right] + (p+1)\left[
            1 +1 +1
        \right]}{(p-1)p(p+2)} 
        = \tfrac{3}{p(p+2)}.
    \end{align*}
    Other moments are derived in a similar fashion. We omit these derivations due to space constraints.

\kern-0.25em
\subsection{Proof of Lemma~\ref{lemma:AKAT}}\label{proof:lemma:AKAT}
\kern-0.125em
    Let the spectral decomposition of $\Kmat$ be denoted as
    $\Kmat = \Lmat \Lambdamat \Lmat\T$,
    where 
    $\Lmat \in \Rbb^{p\times p}$ 
    is an orthogonal matrix,
    and 
    $\Lambdamat = \diag(s_i) \in \Rbb^{p \times p}$, $i = 1,\, \dots,\, p$,
    contains the eigenvalues of $\Kmat$.
    We note that $\Amat\Lmat \sim \Amat$,
    due to the rotational invariance of the standard Gaussian distribution.
    %
    %
    Hence 
    $
        \eunder{\Amat}\big[\Amat\Kmat\Amat\T\big]
        = \eunder{\Amat}\big[\Amat\Lambdamat\Amat\T\big]
        = \eunder{\Amat}\left[\sum_{i=1}^{p} s_i \avec_i \avec_i\T \right],
    $
    where $\avec_i \in \Rbb^{n\times 1}$
    denote the columns of $\Amat$.
    We note that $\avec_i$ are i.i.d. standard Gaussian random vectors,
    hence $\Ebb[\avec_i\avec_i\T] = \eye{n}$. Hence,
    $
        \eunder{\Amat}\big[\Amat\Kmat \Amat\T\big]
        = \left(\sum_{i=1}^p s_i\right) \eye{n}
        = \tr(\Kmat) \eye{n},
    $
    which concludes the proof.

    \kern-0.25em
    
    \bstctlcite{IEEEexample:BSTcontrol}
    \bibliographystyle{IEEEtran}
    \bibliography{ref}    
    
\end{document}